\documentclass[11pt]{article}
\usepackage{amssymb,amsmath,amsfonts}
\usepackage{graphicx}
\usepackage{graphics}
\usepackage{eepic,epsfig}

\@addtoreset{equation}{section}

\makeatother

\begin{document}

\title{Wightman function and vacuum densities for a $Z_{2}$-symmetric thick
brane in AdS spacetime}
\author{A. A. Saharian\thanks{%
E-mail: saharian@ictp.it } \thinspace and A. L. Mkhitaryan \\
\\
\textit{Department of Physics, Yerevan State University} \\
\textit{1 Alex Manoogian Street, 375025 Yerevan, Armenia}}
\maketitle

\begin{abstract}
Positive frequency Wightman function, vacuum expectation values of the field
square and the energy-momentum tensor induced by a $Z_{2}$-symmetric brane
with finite thickness located on $(D+1)$- dimensional AdS background are
evaluated for a massive scalar field with general curvature coupling
parameter. For the general case of static plane symmetric interior structure
the expectation values in the region outside the brane are presented as the
sum of free AdS and brane induced parts. For a conformally coupled massless
scalar the brane induced part in the vacuum energy-momentum tensor vanishes.
In the limit of strong gravitational fields the brane induced parts are
exponentially suppressed for points not too close to the brane boundary. As
an application of general results a special model is considered in which the
geometry inside the brane is a slice of the Minkowski spacetime orbifolded
along the direction perpendicular to the brane. For this model the Wightman
function, vacuum expectation values of the field square and the
energy-momentum tensor inside the brane are evaluated as well and their
behavior is discussed in various asymptotic regions of the parameters. It is
shown that for both minimally and conformally coupled scalar fields the
interior vacuum forces acting on the brane boundaries tend to decrease the
brane thickness.
\end{abstract}

\bigskip


\bigskip

\section{Introduction}

\label{sec:introd}

Recent proposals of large extra dimensions use the concept of brane as a
sub-manifold embedded in a higher dimensional spacetime, on which the
Standard Model particles are confined (see, for instance \cite{Ruba01}).
Braneworlds naturally appear in string/M-theory context and provide a novel
setting for discussing phenomenological and cosmological issues related to
extra dimensions. The models introduced by Randall and Sundrum are
particularly attractive. In RS 2-brane model \cite{Rand99a} the
corresponding background solution consists of two parallel flat 3-branes,
one with positive tension and another with negative tension embedded in a
five dimensional AdS bulk. The fifth coordinate is compactified on $%
S^{1}/Z_{2}$, and the branes are on the two fixed points. It is assumed that
all matter fields are confined on the branes and only the gravity propagates
freely in the five dimensional bulk. In RS 1-brane model \cite{Rand99b}
there is only one positive tension brane in a five dimensional AdS bulk.
This model is obtained from 2-brane one by sending the negative tension
brane to infinity. Due to the warp factor in the AdS metric the infinite
extra dimension gives a finite contribution to the volume. More recently,
alternatives to confining particles on the brane have been investigated and
scenarios with additional bulk fields have been considered.

Motivated by the problems of the radion stabilization and the generation of
cosmological constant, the role of quantum effects in braneworlds has
attracted great deal of attention \cite{Fabi00}-\cite{Saha06c}. A class of
higher dimensional models with compact internal spaces is considered in \cite%
{Flac03b}. Many of treatments of quantum fields in braneworlds
deal mainly with the case of the idealized brane with zero
thickness. This simplification suffers from the disatvantage that
the curvature tensor is singular at the brane location. In
addition, the vacuum expectation values of the local physical
observables diverge on the brane. From a more realistic point of
view we expect that the branes have a finite thickness and the
thickness can act as natural regulator for surface divergences.
The finite core effects also lead to the modification of the
Friedmann equation describing the cosmological evolution inside
the brane. In particular, a phase with accelerated expansion on
the brane can be obtained in the model with pressureless matter as
the only non-gravitational source in the modified Friedmann
equation. In sting theory there exists the minimum length scale
and we cannot neglect the thickness of the corresponding branes at
the string scale. The branes modelled by field theoretical domain
walls have a characteristic thickness determined by the energy
scale where the symmetry of the system is spontaneously broken.
Various models are considered for a thick brane (see, for
instance, \cite{Dewo00} and references therein). Mainly, these
models are constructed as solutions to the coupled Einstein-scalar
equations by choosing a suitable potential for the scalar field.
Vacuum fluctuations for a simple thick de Sitter brane supported
by a bulk scalar field with an axion like potential and the
self-consistency of this braneworld are investigated in Refs.
\cite{Mina06a}.

In the present paper we investigate the effects of core on
properties of the quantum vacuum for the general plane symmetric
static model of the brane with finite thickness. The most
important quantities characterizing these properties are the
vacuum expectation values of the field square and the
energy-momentum tensor. Though the corresponding operators are
local, due to the global nature of the vacuum, the vacuum
expectation values describe the global properties of the bulk and
carry an important information about the internal structure of the
brane. In addition to describing the physical structure of the
quantum field at a given point, the energy-momentum tensor acts as
the source of gravity in the Einstein equations. It therefore
plays an important role in modelling a self-consistent dynamics
involving the gravitational field. As the first step for the
investigation of vacuum densities we evaluate the positive
frequency Wightman function for a massive scalar field with
general curvature coupling parameter. This function gives
comprehensive insight into vacuum fluctuations and determines the
response of a particle detector of the Unruh-DeWitt type moving in
the brane bulk. The problem under consideration is also of
separate interest as an example with gravitational and
boundary-induced polarizations of the vacuum, where all
calculations can be performed in a closed form. The corresponding
results specify the conditions under which we can ignore the
details of the interior structure and approximate the effect of
the brane by the idealized model. In addition, as it will be shown
below, the phenomenological parameters in the zero-thickness brane
models such as brane mass terms for scalar fields are calculable
in terms of the inner structure of the brane  within the framework
of the model considered in the present paper.

The plan of this paper is as follows. In Section \ref{sec:WF} we consider
the Wightman function in the exterior of the brane for the general structure
of the core assuming that the components of the metric tensor and their
derivatives are continuous at the transition surface between the core and
the exterior. The Section \ref{sec:thinshell} is devoted to the
generalization of the corresponding results when an additional surface shell
is present on the bounding surfaces between the core and the exterior. By
using the formulae for the Wightman function, in Section \ref{sec:Outside}
we investigate the vacuum expectation values of the field square and the
energy-momentum tensor. As an illustration of the general results, in
Section \ref{sec:flowerpot} we consider a model with Minkowskian geometry
inside the brane. For this model the vacuum expectation values inside the
core are investigated as well. The last section contains a summary of the
work.

\section{Wightman function}

\label{sec:WF}

\subsection{Model}

We consider a domain wall with finite thickness on background of $(D+1)$%
-dimensional AdS spacetime. As in RS 1-brane scenario we will assume that
our model is $Z_{2}$-symmetric with respect to the plane $y=0$ located at
the center of the brane. The spacetime is described by two distinct metric
tensors in the regions outside and inside the brane. The corresponding line
element in the exterior region $|y|>a$ has the form
\begin{equation}
ds^{2}=g_{ik}dx^{i}dx^{k}=e^{-2k_{D}|y|}\eta _{\mu \nu }dx^{\mu }dx^{\nu
}-dy^{2},  \label{metric}
\end{equation}%
where $\eta _{\mu \nu }=\mathrm{diag}(1,-1,\ldots ,-1)$ is the metric for
the $D $-dimensional Minkowski spacetime, $1/k_{D}$ is the AdS curvature
radius. Here and below $i,k=0,1,\ldots ,D$, and $\mu ,\nu =0,1,\ldots ,D-1$.
We will assume that inside the brane (region $|y|<a$) the spacetime geometry
is regular and is described by the static plane symmetric line element%
\begin{equation}
ds^{2}=e^{2u(y)}dt^{2}-e^{2v(y)}d\mathbf{x}^{2}-e^{2w(y)}dy^{2},
\label{metricinside}
\end{equation}%
where $\mathbf{x}=(x^{1},\ldots ,x^{D-1})$ are coordinates parallel to the
brane. Due to the $Z_{2}$-symmetry the functions $u(y)$, $v(y)$, $w(y)$ are
even functions on $y$. These functions are continuous at the core boundary:
\begin{equation}
u(a)=v(a)=-k_{D}a,\;w(a)=0.  \label{uvbound}
\end{equation}%
Here we assume that there is no surface energy-momentum tensor located at $%
|y|=a$ and, hence, the derivatives of these functions are continuous as
well. The generalization to the case with an infinitely thin shell at the
boundary of two metrics will be discussed in the next section.

For the metric corresponding to line element (\ref{metricinside}) the
nonzero components of the Ricci tensor are given by expressions (no
summation over $i$)%
\begin{eqnarray}
R_{0}^{0} &=&-e^{-2w}\left[ u^{\prime \prime }+u^{\prime 2}+(D-1)u^{\prime
}v^{\prime }-u^{\prime }w^{\prime }\right] ,  \notag \\
R_{i}^{i} &=&-e^{-2w}\left[ v^{\prime \prime }+(D-1)v^{\prime 2}+v^{\prime
}u^{\prime }-v^{\prime }w^{\prime }\right] ,\;i=1,\ldots ,D-1,
\label{Riccin} \\
R_{D}^{D} &=&-e^{-2w}\left[ u^{\prime \prime }+(D-1)v^{\prime \prime
}+u^{\prime 2}+(D-1)v^{\prime 2}-w^{\prime }u^{\prime }-(D-1)w^{\prime
}v^{\prime }\right] ,  \notag
\end{eqnarray}%
where the prime means the derivative with respect to the coordinate $y$. The
corresponding Ricci scalar has the form%
\begin{eqnarray}
R &=&-2e^{-2w}\left[ u^{\prime \prime }+(D-1)v^{\prime \prime }+u^{\prime
2}+D(D-1)v^{\prime 2}/2\right.  \notag \\
&&\left. +(D-1)u^{\prime }v^{\prime }-u^{\prime }w^{\prime }-(D-1)v^{\prime
}w^{\prime }\right] .  \label{Richscin}
\end{eqnarray}%
In the special case $u(y)=v(y)$ one has $R_{0}^{0}=R_{i}^{i}$, $i=1,\ldots
,D-1$, and the model is Poincare invariant along the directions parallel to
the brane.

\subsection{Eigenfunctions}

In this paper we are interested in the vacuum polarization effects for a
scalar field with general curvature coupling parameter $\xi $ propagating in
the bulk described above. The corresponding field equation has the form%
\begin{equation}
\left( \nabla _{i}\nabla ^{i}+m^{2}+\xi R\right) \varphi =0,\;\nabla
_{i}\nabla ^{i}=\frac{1}{\sqrt{|g|}}\partial _{i}(g^{ik}\sqrt{|g|}\partial
_{k}),  \label{fieldeq}
\end{equation}%
where $\nabla _{i}$ is the covariant derivative operator associated with
line element (\ref{metric}) outside the brane and with line element (\ref%
{metricinside}) inside the brane. The values of the curvature coupling
parameter $\xi =0$ and $\xi =\xi _{D}$ with $\xi _{D}\equiv (D-1)/4D$
correspond to the most important special cases of minimally and conformally
coupled scalar fields. As a first stage for the evaluation of the vacuum
expectation values (VEVs) for the field square and the energy-momentum
tensor we consider the positive frequency Wightman function $\langle
0|\varphi (x)\varphi (x^{\prime })|0\rangle $, where $|0\rangle $ is the
amplitude for the corresponding vacuum state. This function also determines
the response of the Unruh-DeWitt type particle detector at a given state of
motion (see, for instance, \cite{Birrell}). The Wightman function can be
evaluated by using the mode sum formula
\begin{equation}
\langle 0|\varphi (x)\varphi (x^{\prime })|0\rangle =\sum_{\alpha }\varphi
_{\alpha }(x)\varphi _{\alpha }^{\ast }(x^{\prime }),  \label{mfieldmodesum}
\end{equation}%
where $\left\{ \varphi _{\alpha }(x),\varphi _{\alpha }^{\ast }(x^{\prime
})\right\} $ is a complete orthonormalized set of positive and negative
frequency solutions to the field equation. The collective index $\alpha $
can contain both discrete and continuous components. In Eq. (\ref%
{mfieldmodesum}) it is assumed summation over discrete indices and
integration over continuous ones.

On the base of the plane symmetry of the problem under consideration the
corresponding eigenfunctions can be presented in the form
\begin{equation}
\varphi _{\alpha }(x^{i})=\phi _{\mathbf{k}}(x^{\mu })f_{\lambda }(y),
\label{eigfunc1}
\end{equation}%
where $\phi _{\mathbf{k}}(x^{\mu })$ are the standard Minkowskian modes on
the hyperplane parallel to the brane:
\begin{equation}
\phi _{\mathbf{k}}(x^{\mu })=\frac{e^{i\mathbf{k}\cdot \mathbf{x}-i\omega t}%
}{\sqrt{2\omega (2\pi )^{D-1}}},\quad \omega =\sqrt{k^{2}+\lambda ^{2}}%
,\quad k=|\mathbf{k}|,  \label{branefunc1}
\end{equation}%
$\lambda $ is the separation constant. Below we will assume that $y\geqslant
0$. The corresponding formulae in the region $y<0$ are obtained from the $%
Z_{2}$-symmetry of the model. Substituting eigenfunctions (\ref{eigfunc1})
into field equation (\ref{fieldeq}), for the function $f(y)$ one obtains the
following equation
\begin{equation}
e^{-u-(D-1)v-w}\partial _{y}\left[ e^{u+(D-1)v-w}\partial _{y}f_{\lambda }%
\right] -\left[ m^{2}+\xi R+k^{2}(e^{-2v}-e^{-2u})-\lambda ^{2}e^{-2u}\right]
f_{\lambda }=0.  \label{eqforfn}
\end{equation}%
For the exterior AdS geometry one has $u(y)=v(y)=-k_{D}y$, and $%
R=-D(D+1)k_{D}^{2}$. In this case the solution to equation (\ref{eqforfn})
for the region $y>a$ is
\begin{equation}
f_{\lambda }(y)=e^{Dk_{D}y/2}\left[ A_{\nu }J_{\nu }(\lambda z)+B_{\nu
}Y_{\nu }(\lambda z)\right] ,\;z=e^{k_{D}y}/k_{D},  \label{fny}
\end{equation}%
where $A_{\nu }$ and $B_{\nu }$ are integration constants, $J_{\nu }(x)$, $%
Y_{\nu }(x)$ are the Bessel and Neumann functions, and we use the notation
\begin{equation}
\nu =\sqrt{D^{2}/4-D(D+1)\xi +m^{2}/k_{D}^{2}}.  \label{nu}
\end{equation}%
Here we will assume values of the curvature coupling parameter for which $%
\nu $ is real. For imaginary $\nu $ the ground state becomes unstable \cite%
{Brei82}. Note that for a conformally coupled massless scalar one has $\nu
=1/2$ and the cylinder functions in Eq. (\ref{fny}) are expressed via the
elementary functions. The regular solution of equation (\ref{eqforfn}) in
the region $y<a$ we will denote by $R(y,k,\lambda )$. Note that the
parameters $k$ and $\lambda $ enter in the radial equation in the form $%
k^{2} $ and $\lambda ^{2}$. As a result the regular solution can be chosen
in such a way that $R(y,-k,\lambda )=\mathrm{const}\cdot R(y,k,\lambda )$
and $R(y,k,-\lambda )=\mathrm{const}\cdot R(y,k,\lambda )$. Now for the
radial part of the eigenfunctions one has%
\begin{equation}
f_{\lambda }(y)=\left\{
\begin{array}{ll}
R(y,k,\lambda ), & \mathrm{for}\;y<a \\
e^{Dk_{D}y/2}\left[ A_{\nu }J_{\nu }(\lambda z)+B_{\nu }Y_{\nu }(\lambda z)%
\right] , & \mathrm{for}\;y>a%
\end{array}%
.\right.  \label{fl}
\end{equation}%
The coefficients $A_{\nu }$ and $B_{\nu }$ are determined by the conditions
of continuity of the radial function and its derivative at $y=a$. From these
conditions we find%
\begin{eqnarray}
R_{a} &=&e^{Dk_{D}a/2}\left[ A_{\nu }J_{\nu }(\lambda z_{a})+B_{\nu }Y_{\nu
}(\lambda z_{a})\right] ,  \label{Ra1} \\
R_{a}^{\prime } &=&k_{D}e^{Dk_{D}a/2}\left\{ A_{\nu }\left[ DJ_{\nu
}(\lambda z_{a})/2+\lambda z_{a}J_{\nu }^{\prime }(\lambda z_{a})\right]
\right.  \notag \\
&&\left. +B_{\nu }\left[ DY_{\nu }(\lambda z_{a})/2+\lambda z_{a}Y_{\nu
}^{\prime }(\lambda z_{a})\right] \right\} ,  \label{Ra2}
\end{eqnarray}%
where%
\begin{equation}
R_{a}=R(a,k,\lambda ),\;R_{a}^{\prime }=\partial _{y}R(y,k,\lambda )|_{y=a}.
\label{Ra12}
\end{equation}%
Solving these equations with respect to $A_{\nu }$ and $B_{\nu }$ and using
the Wronskian for the Bessel and Neumann functions, one has%
\begin{equation}
A_{\nu }=\frac{\pi }{2}e^{-Dk_{D}a/2}R_{a}\bar{Y}_{\nu }(\lambda
z_{a}),\;B_{\nu }=-\frac{\pi }{2}e^{-Dk_{D}a/2}R_{a}\bar{J}_{\nu }(\lambda
z_{a}).  \label{AnuBnu}
\end{equation}%
Here and in what follows for a cylinder function $F(z)$ we use the notation
\begin{equation}
\bar{F}(z)\equiv zF^{\prime }(z)+\left( \frac{D}{2}-\frac{R_{a}^{\prime }}{%
k_{D}R_{a}}\right) F(z).  \label{barnot}
\end{equation}%
Note that due to our choice of the function $R(y,k,\lambda )$, the
logarithmic derivative in formula (\ref{barnot}) is an even function on $%
\lambda $. Hence, in the region $y>a$ the radial part of the eigenfunctions
has the form%
\begin{equation}
f_{\lambda }(y)=\frac{\pi }{2}e^{Dk_{D}(y-a)/2}R_{a}g_{\nu }(\lambda
z_{a},\lambda z),  \label{fl2}
\end{equation}%
where the notation
\begin{equation}
g_{\nu }(u,v)=\bar{Y}_{\nu }(u)J_{\nu }(v)-\bar{J}_{\nu }(u)Y_{\nu }(v)
\label{gnot}
\end{equation}%
is introduced.

For the eigenfunctions we have the following orthonormalization condition%
\begin{equation}
\int dyd^{D-1}x\sqrt{|g|}g^{00}\varphi _{\alpha }(x)\varphi _{\alpha
^{\prime }}^{\ast }(x)=\frac{\delta _{\alpha \alpha ^{\prime }}}{2\omega },
\label{ortnorm}
\end{equation}%
where $\delta _{\alpha \alpha ^{\prime }}$ is understood as the Kronecker
symbol for discrete indices and as the Dirac delta function for continuous
ones. Substituting eigenfunctions (\ref{eigfunc1}), the normalization
condition is written in terms of the radial eigenfunctions
\begin{equation}
\int_{-\infty }^{+\infty }dy\sqrt{|g|}g^{00}f_{\lambda }(y)f_{\lambda
^{\prime }}(y)=\delta (\lambda -\lambda ^{\prime }).  \label{ortcond}
\end{equation}%
As the integral on the left is divergent for $\lambda ^{\prime }=\lambda $,
the main contribution in the coincidence limit comes from large values $|y|$%
. \ By using the expression (\ref{fl2}) for the radial part in the region $%
|y|>a$ and replacing the Bessel and Neumann functions by the leading terms
of their asymptotic expansions for large values of the argument, we obtain
the following result:%
\begin{equation}
R_{a}^{-2}=\frac{\pi ^{2}}{2}\frac{\bar{J}_{\nu }^{2}(\lambda z_{a})+\bar{Y}%
_{\nu }^{2}(\lambda z_{a})}{z_{a}^{D}k_{D}^{D-1}\lambda }.
\label{normcoefRl}
\end{equation}%
This relation determines the normalization coefficient for the interior
eigenfunctions.

In addition to the modes with real $\lambda $ discussed above, modes with
purely imaginary $\lambda $ can exist. For this type of modes the radial
eigenfunctions in the region $y>a$ are given by the expression $C_{\nu
}e^{Dk_{D}y/2}K_{\nu }(|\lambda |z)$, where $K_{\nu }(x)$ is the MacDonald
function. From the continuity of the eigenfunctions at $y=a$ one finds $%
R_{a}=C_{\nu }e^{Dk_{D}a/2}K_{\nu }(|\lambda |z_{a})$, and from the
continuity of the radial derivative we see that $|\lambda |$ is a solution
of the equation%
\begin{equation}
|\lambda |z_{a}\frac{K_{\nu }^{\prime }(|\lambda |z_{a})}{K_{\nu }(|\lambda
|z)}+\frac{D}{2}=\frac{\partial _{y}R(y,k,e^{\pi i/2}|\lambda |)|_{y=a}}{%
k_{D}R(a,k,e^{\pi i/2}|\lambda |)}.  \label{immodes}
\end{equation}
As it follows from (\ref{branefunc1}), for the modes with purely imaginary $%
\lambda $ the corresponding frequency is imaginary in the region $k<|\lambda
|$ and the vacuum state becomes unstable. To have a stable ground state, in
the discussion below we assume that there are no modes with imaginary $%
\lambda $.

\subsection{Wightman function in the exterior region}

Substituting the eigenfunctions (\ref{eigfunc1}) into the mode sum (\ref%
{mfieldmodesum}), for the Wightman function one finds%
\begin{eqnarray}
\langle 0|\varphi (x)\varphi (x^{\prime })|0\rangle &=&\frac{k_{D}^{D-1}}{4}%
(zz^{\prime })^{\frac{D}{2}}\int d\mathbf{k}\frac{e^{i\mathbf{k}\cdot (%
\mathbf{x}-\mathbf{x}^{\prime })}}{(2\pi )^{D-1}}\int_{0}^{\infty }d\lambda
\frac{\lambda \,}{\sqrt{k^{2}+\lambda ^{2}}}  \notag \\
&&\times \frac{g_{\nu }(\lambda z_{a},\lambda z)g_{\nu }(\lambda
z_{a},\lambda z^{\prime })}{\bar{J}_{\nu }^{2}(\lambda z_{a})+\bar{Y}_{\nu
}^{2}(\lambda z_{a})}e^{i(t^{\prime }-t)\sqrt{k^{2}+\lambda ^{2}}}.
\label{Wf2}
\end{eqnarray}%
For further transformation of this formula we use the identity
\begin{equation}
\frac{g_{\nu }(\lambda z_{a},\lambda z)g_{\nu }(\lambda z_{a},\lambda
z^{\prime })}{\bar{J}_{\nu }^{2}(\lambda z_{a})+\bar{Y}_{\nu }^{2}(\lambda
z_{a})}=J_{\nu }(\lambda z)J_{\nu }(\lambda z^{\prime })-\frac{1}{2}%
\sum_{s=1,2}\frac{\bar{J}_{\nu }(\lambda z_{a})}{\bar{H}_{\nu
}^{(s)}(\lambda z_{a})}H_{\nu }^{(s)}(\lambda z)H_{\nu }^{(s)}(\lambda
z^{\prime }),  \label{ident1}
\end{equation}%
where $H_{\nu }^{(s)}(x)$, $s=1,2$ are the Hankel functions. This allows us
to present the Wightman function in the form
\begin{equation}
\langle 0|\varphi (x)\varphi (x^{\prime })|0\rangle =\frac{1}{2}\langle
0_{S}|\varphi (x)\varphi (x^{\prime })|0_{S}\rangle +\langle \varphi
(x)\varphi (x^{\prime })\rangle _{\mathrm{b}},  \label{coreWF}
\end{equation}%
where
\begin{eqnarray}
\langle 0_{S}|\varphi (x)\varphi (x^{\prime })|0_{S}\rangle &=&\frac{%
k_{D}^{D-1}}{2}(zz^{\prime })^{\frac{D}{2}}\int d\mathbf{k}\frac{e^{i\mathbf{%
k}\cdot (\mathbf{x}-\mathbf{x}^{\prime })}}{(2\pi )^{D-1}}\int_{0}^{\infty
}d\lambda \frac{\lambda \,}{\sqrt{k^{2}+\lambda ^{2}}}  \notag \\
&&\times J_{\nu }(\lambda z)J_{\nu }(\lambda z^{\prime })e^{i(t^{\prime }-t)%
\sqrt{k^{2}+\lambda ^{2}}},  \label{Mink}
\end{eqnarray}%
is the positive frequency Wightman function for the AdS spacetime without
boundaries (see, for instance, \cite{Saha05}), and the part
\begin{eqnarray}
\langle \varphi (x)\varphi (x^{\prime })\rangle _{\mathrm{b}} &=&-\frac{%
k_{D}^{D-1}}{8}(zz^{\prime })^{\frac{D}{2}}\int d\mathbf{k}\frac{e^{i\mathbf{%
k}\cdot (\mathbf{x}-\mathbf{x}^{\prime })}}{(2\pi )^{D-1}}%
\sum_{s=1}^{2}\int_{0}^{\infty }d\lambda \,\lambda  \notag \\
&&\times \frac{e^{i\sqrt{\lambda ^{2}+k^{2}}(t^{\prime }-t)}}{\sqrt{\lambda
^{2}+k^{2}}}\frac{\bar{J}_{\nu }(\lambda z_{a})}{\bar{H}_{\nu
}^{(s)}(\lambda z_{a})}H_{\nu }^{(s)}(\lambda z)H_{\nu }^{(s)}(\lambda
z^{\prime }),  \label{extdif}
\end{eqnarray}%
induced by the brane. Quantum effects in free AdS spacetime are well
investigated in literature (see references given in \cite{Saha05}) and in
the discussion below we will be mainly concentrated on the effects induced
by the brane.

On the right of formula (\ref{extdif}) we rotate the integration contour in
the complex plane $\lambda $ by the angle $\pi /2$ for $s=1$ and by the
angle $-\pi /2$ for $s=2$. Under the condition $z+z^{\prime
}>2z_{a}+|t-t^{\prime }|$, the contributions from the integrals over the
arcs of the semicircle with large radius in the upper/lower half-plane tend
to zero. In particular, this condition is satisfied in the coincidence limit
for points outside the brane. Further, by using the property that the
logarithmic derivative of the function $R(y,k,\lambda )$ in formula (\ref%
{barnot}) is an even function on $\lambda $, we can see that the integrals
over the segments $(0,ik)$ and $(0,-ik)$ of the imaginary axis cancel out.
As a result, after introducing the modified Bessel functions, the
brane-induced part can be presented in the form
\begin{eqnarray}
\langle \varphi (x)\varphi (x^{\prime })\rangle _{\mathrm{b}} &=&-\frac{%
k_{D}^{D-1}}{(2\pi )^{D}}(zz^{\prime })^{\frac{D}{2}}\int d\mathbf{k}\,e^{i%
\mathbf{k}\cdot (\mathbf{x}-\mathbf{x}^{\prime })}\int_{k}^{\infty }d\lambda
\lambda \frac{\tilde{I}_{\nu }(\lambda z_{a})}{\tilde{K}_{\nu }(\lambda
z_{a})}  \notag \\
&&\times \frac{K_{\nu }(\lambda z)K_{\nu }(\lambda z^{\prime })}{\sqrt{%
\lambda ^{2}-k^{2}}}\cosh \!\left[ \sqrt{\lambda ^{2}-k^{2}}(t^{\prime }-t)%
\right] .  \label{regWightout}
\end{eqnarray}%
Here and below the tilted notation for the modified Bessel functions $I_{\nu
}(x)$ and $K_{\nu }(x)$ is defined by the formula%
\begin{equation}
\tilde{F}(x)\equiv xF^{\prime }(x)+\mathcal{R}(a,k,x)F(x),
\label{Barrednotmod}
\end{equation}%
with the notation%
\begin{equation}
\mathcal{R}(a,k,x)=\frac{D}{2}-\frac{\partial _{y}R(y,k,xe^{\pi
i/2}/z_{a})|_{y=a}}{k_{D}R(a,k,xe^{\pi i/2}/z_{a})}.  \label{Rlcal}
\end{equation}%
Note that, by taking into account this notation, equation (\ref{immodes})
for the modes with imaginary $\lambda $ is written in the form $\tilde{K}%
_{\nu }(|\lambda |z_{a})=0$.

As we see from (\ref{regWightout}), the information about the inner
structure of the brane is contained in the logarithmic derivative of the
interior radial function in formula (\ref{Rlcal}). In formula (\ref%
{regWightout}) the integration over the angular part can be done by using
the formula
\begin{equation}
\int d\mathbf{k}\,\frac{e^{i\mathbf{k}\mathbf{x}}F(k)}{(2\pi )^{(D-1)/2}}%
=\int_{0}^{\infty }dk\,k^{D-2}F(k)\frac{J_{(D-3)/2}(k|\mathbf{x}|)}{(k|%
\mathbf{x}|)^{(D-3)/2}},  \label{angint}
\end{equation}%
for a given function $F(k)$. In the RS 1-brane model with the brane of zero
thickness the brane induced part in the Wightman function is given by a
formula similar to (\ref{regWightout}) with the replacement \cite{Saha05}
\begin{equation}
\mathcal{R}(a,k,x)\rightarrow D/2-2D\xi -c/2k_{D},  \label{replaceR}
\end{equation}%
in the definition (\ref{Barrednotmod}) of the tilted notation. On
the right of relation (\ref{replaceR}), the parameter $c$ is the
brane mass term for a scalar field which is a phenomenological
parameter in the model with zero thickness brane. As we see, in
the model under consideration the effective brane mass term is
determined by the inner structure of the core. Note that in RS
2-brane model the mass terms on separate branes determine the
1-loop effective potential for the radion field and play an
important role in the stabilization of the interbrane distance.

\section{Model with an infinitely thin shell on the brane boundary}

\label{sec:thinshell}

The results considered in the previous section can be generalized to the
models where an additional infinitely thin plane shell located at $y=a$ is
present with the surface energy-momentum tensor $\tau _{i}^{k}$. Let $n^{i}$
be the normal to the shell normalized by the condition $n_{i}n^{i}=-1$. We
assume that it points into the bulk on both sides of the shell. From the
Israel matching conditions one has%
\begin{equation}
\left\{ K_{ik}-Kh_{ik}\right\} =8\pi G\tau _{ik},  \label{matchcond}
\end{equation}%
where the curly brackets denote summation over each side of the shell, $%
h_{ik}=g_{ik}+n_{i}n_{k}$ is the induced metric on the shell, $%
K_{ik}=h_{i}^{r}h_{k}^{s}\nabla _{r}n_{s}$ its extrinsic curvature and $%
K=K_{i}^{i}$. For the shell located on the boundary $y=a$ and for the region
$y\leqslant a$ one has $n_{i}=\delta _{i}^{D}e^{w(y)}$ and the non-zero
components of the extrinsic curvature are given by the formulae%
\begin{equation}
K_{0}^{0}=-u^{\prime }(a-),\;K_{i}^{j}=-\delta _{i}^{j}v^{\prime
}(a-),\;i=1,2,,\ldots ,D-1.  \label{exttensor}
\end{equation}%
The corresponding expressions for the region $y\geqslant a$ are obtained by
taking $u(y)=v(y)=-k_{D}y$, $w(y)=0$ and changing the signs for the
components of the extrinsic curvature tensor:%
\begin{equation}
K_{i}^{j}=-\delta _{i}^{j}k_{D},\;i=1,2,,\ldots ,D-1,\;y=a+.
\label{exttensorout}
\end{equation}%
Now from the matching conditions (\ref{matchcond}) we find (no summation
over $i$)%
\begin{eqnarray}
u^{\prime }(a-) &=&-k_{D}+8\pi G\left( \tau _{i}^{i}-\frac{D-2}{D-1}\tau
_{0}^{0}\right) ,\;i=1,2,\ldots ,D-1,  \label{matchcond1} \\
v^{\prime }(a-) &=&-k_{D}+\frac{8\pi G}{D-1}\tau _{0}^{0},
\label{matchcond2}
\end{eqnarray}%
where $f^{\prime }(a-)$ is understood in the sense $\lim_{y\rightarrow
a-0}f^{\prime }(y)$. For models with Poincare invariance along the
directions parallel to the brane one has $\tau _{i}^{i}=\tau _{0}^{0}$.

The discontinuity of the functions $u^{\prime }(y)$ and $v^{\prime }(y)$ at $%
y=a$ leads to the delta function term%
\begin{equation}
2\left[ u^{\prime }(a-)+(D-1)v^{\prime }(a-)+Dk_{D}\right] \delta (y-a)
\label{deltain Ricci}
\end{equation}%
in the Ricci scalar and, hence, in the equation (\ref{eqforfn}) for the
radial eigenfunctions. Note that the expression in the square brackets is
related to the surface energy-momentum tensor by the formula%
\begin{equation}
u^{\prime }(a-)+(D-1)v^{\prime }(a-)+Dk_{D}=\frac{8\pi G}{D-1}\tau ,
\label{Tracesurf}
\end{equation}%
where $\tau $ is the trace of the surface energy-momentum tensor.

For a non-minimally coupled scalar field, due to the delta function term in
the equation for the radial eigenfunctions, these functions have a
discontinuity in their slope at $y=a$. The corresponding jump condition is
obtained by integrating the equation (\ref{eqforfn}) through the point $y=a$:%
\begin{equation}
f_{\lambda }^{\prime }(a+)-f_{\lambda }^{\prime }(a-)=\frac{16\pi G\xi }{D-1}%
\tau f_{\lambda }(a).  \label{flderjump}
\end{equation}%
Now the coefficients in the formulae (\ref{fl}) for the eigenfunctions are
determined by the continuity condition for the radial eigenfunctions and by
the jump condition for their radial derivative. It can be seen that the
corresponding eigenfunctions are given by the same formulae (\ref{fl2}) and (%
\ref{normcoefRl}) with the new barred notation%
\begin{equation}
\bar{F}(z)\equiv zF^{\prime }(z)+\left( \frac{D}{2}-\frac{16\pi G\xi }{%
(D-1)k_{D}}\tau -\frac{R_{a}^{\prime }}{k_{D}R_{a}}\right) F(z).
\label{barrednew}
\end{equation}%
Hence, the part in the exterior Wightman function induced by the brane is
given by formula (\ref{regWightout}), where the tilted notation is defined
by Eq. (\ref{Barrednotmod}) with the function%
\begin{equation}
\mathcal{R}(a,k,x)=\frac{D}{2}-\frac{16\pi G\xi }{(D-1)k_{D}}\tau -\frac{%
\partial _{y}R(y,k,xe^{\pi i/2}/z_{a})|_{y=a}}{k_{D}R(a,k,xe^{\pi i/2}/z_{a})%
}.  \label{newRlcal}
\end{equation}%
The trace of the surface energy-momentum tensor in this expression is
related to the components of the metric tensor inside the brane by formula (%
\ref{Tracesurf}). As before, in the model with surface energy-momentum
tensor the equation for the modes with imaginary $\lambda $ is written in
the form $\tilde{K}_{\nu }(|\lambda |z_{a})=0$.

\section{Vacuum expectation values outside the brane}

\label{sec:Outside}

\subsection{VEV of the field square}

The VEV of the field square is obtained by computing the Wightman function
in the coincidence limit $x^{\prime }\rightarrow x$. In this limit
expression (\ref{coreWF}) gives a divergent result and some renormalization
procedure is needed. Outside the brane the local geometry is the same as
that for the AdS spacetime. Hence, in the region $y>a$ the renormalization
procedure for the local characteristics of the vacuum, such as the field
square and the energy-momentum tensor, is the same as for the free AdS
spacetime.

By using the formula for the Wightman function from the previous section,
the VEV of the field square in the exterior region is presented in the form%
\begin{equation}
\langle 0|\varphi ^{2}|0\rangle =\frac{1}{2}\langle 0_{S}|\varphi
^{2}|0_{S}\rangle +\langle \varphi ^{2}\rangle _{\mathrm{b}},
\label{phi2ext}
\end{equation}%
where $\langle 0_{S}|\varphi ^{2}|0_{S}\rangle $ is the VEV of the field
square in the free AdS spacetime. The part induced by the brane is directly
obtained from formula (\ref{regWightout}) in the coincidence limit:%
\begin{equation}
\langle \varphi ^{2}\rangle _{\mathrm{b}}=-\frac{\pi
^{-(D+1)/2}k_{D}^{D-1}z^{D}}{2^{D-1}\Gamma ((D-1)/2)}\int_{0}^{\infty
}dk\,k^{D-2}\,\int_{k}^{\infty }d\lambda \,\lambda \frac{\tilde{I}_{\nu
}(\lambda z_{a})}{\tilde{K}_{\nu }(\lambda z_{a})}\frac{K_{\nu }^{2}(\lambda
z)}{\sqrt{\lambda ^{2}-k^{2}}}.  \label{phi2cext0}
\end{equation}%
The VEV\ of the field square in the free AdS spacetime is well investigated
in literature \cite{Burg85} and is given by the formula
\begin{equation}
\langle 0_{S}|\varphi ^{2}|0_{S}\rangle =\frac{k_{D}^{D-1}}{(4\pi )^{(D+1)/2}%
}\frac{\Gamma \left( \frac{1-D}{2}\right) \Gamma \left( \frac{D}{2}+\nu
\right) }{\Gamma (1+\nu -D/2)}.  \label{phi2AdS}
\end{equation}%
This VEV does not depend on the spacetime point, which is a direct
consequence of the maximal symmetry of the AdS bulk. For even $D$, the
expression on the right of (\ref{phi2AdS}) is finite and can be taken as the
renormalized value for the field square. For odd $D$, an additional
renormalization removing the pole of the gamma function in the numerator is
necessary.

By taking into account that the tilted functions in formula (\ref{phi2cext0}%
) depend on $k $, in the general case of the core model the corresponding
expression cannot be further simplified. This can be done in the model with
Poincare invariance along the directions parallel to the brane. For this
special case in (\ref{metricinside}) we have $u(y)=v(y)$ and the term with $%
k^{2}$ in (\ref{eqforfn}) disappears. As a result the interior radial
functions $R(y,k,\lambda )$ do not depend on $k$, $R(y,k,\lambda
)=R(y,\lambda )$, and, hence the same is the case for the tilted functions
in (\ref{phi2cext0}). In this case the formula for the VEV of the field
square is further simplified by using the formula%
\begin{equation}
\int_{0}^{\infty }dkk^{D-2}\int_{k}^{\infty }\frac{f(\lambda )d\lambda }{%
\sqrt{\lambda ^{2}-k^{2}}}=\frac{\sqrt{\pi }\Gamma \left( \frac{D-1}{2}%
\right) }{2\Gamma \left( \frac{D}{2}\right) }\int_{0}^{\infty }dxx^{D-2}f(x).
\label{Hash1}
\end{equation}%
This leads to the result%
\begin{equation}
\langle \varphi ^{2}\rangle _{\mathrm{b}}=-\frac{k_{D}^{D-1}z^{D}}{(4\pi
)^{D/2}\Gamma \left( D/2\right) }\int_{0}^{\infty }dxx^{D-1}\,\frac{\tilde{I}%
_{\nu }(xz_{a})}{\tilde{K}_{\nu }(xz_{a})}K_{\nu }^{2}(xz),  \label{phi2cext}
\end{equation}%
where the tilted notations are defined by (\ref{Barrednotmod}).

The integral in Eq. (\ref{phi2cext}) is exponentially convergent in the
upper limit for $z>z_{a}$ and diverges for points on the boundary of the
brane. The surface divergences in the VEVs of local physical observables
bilinear in the field are well known in quantum field theory with
boundaries. They can be regularized by considering more realistic model with
smooth transition between the interior and exterior metrics. By taking into
account that for a given $z$ the main contribution into the integral in (\ref%
{phi2cext}) comes from the region $x\lesssim (z-z_{a})^{-1}$, we expect that
under the condition $y-a\gg y_{0}$, with $y_{0}$ being the thickness of the
transition range, the results of the present paper will be valid for this
kind of models as well.

At large distances from the brane, $z\gg z_{a}$, we introduce a new
integration variable $y=xz$ and expand the integrand over $z_{a}/z$. By
using the formula for the integral involving the square of the MacDonald
function, to the leading order we obtain%
\begin{equation}
\langle \varphi ^{2}\rangle _{\mathrm{b}}=-\frac{k_{D}^{D-1}(z_{a}/z)^{2\nu }%
}{2^{D+2\nu +1}\pi ^{(D-1)/2}}\frac{\mathcal{R}(a)+\nu }{\mathcal{R}(a)-\nu }%
\frac{\Gamma (D/2+\nu )\Gamma (D/2+2\nu )}{\nu \Gamma ^{2}(\nu )\Gamma
((D+1)/2+\nu )},  \label{phi2cextfar}
\end{equation}%
where we have introduced the notation
\begin{equation}
\mathcal{R}(a)=\frac{D}{2}-\frac{16\pi G\xi }{(D-1)k_{D}}\tau -\frac{%
\partial _{y}R(y,0)|_{y=a}}{k_{D}R(a,0)}.  \label{Rcal0}
\end{equation}%
As we see, at large distances from the brane the brane induced part is
exponentially suppressed by the factor $\exp (-2\nu k_{D}y)$.

\subsection{Vacuum energy-momentum tensor}

Now we turn to the investigation of the VEV of the energy-momentum tensor in
the region $y>a$. Having the Wightman function and the VEV\ for the field
square, this can be done on the base of the formula
\begin{equation}
\langle 0|T_{ik}|0\rangle =\lim_{x^{\prime }\rightarrow x}\partial
_{i}\partial _{k}^{\prime }\langle 0|\varphi (x)\varphi (x^{\prime
})|0\rangle +\left[ \left( \xi -\frac{1}{4}\right) g_{ik}\nabla _{l}\nabla
^{l}-\xi \nabla _{i}\nabla _{k}-\xi R_{ik}\right] \langle 0|\varphi
^{2}|0\rangle .  \label{mvevEMT}
\end{equation}%
Note that on the left of this formula we have used the expression for the
energy-momentum tensor of a scalar field which differs from the standard one
by the term which vanishes on the solutions of the field equation (\ref%
{fieldeq}) (see Ref. \cite{SahaSurf}). Similar to the Wightman function, the
components of the vacuum energy-momentum tensor are presented in the
decomposed form%
\begin{equation}
\langle 0|T_{ik}|0\rangle =\frac{1}{2}\langle 0_{S}|T_{ik}|0_{S}\rangle
+\langle T_{ik}\rangle _{\mathrm{b}},  \label{Tikextdecomp}
\end{equation}%
where $\langle 0_{S}|T_{ik}|0_{S}\rangle $ is the vacuum energy-momentum
tensor in the free AdS spacetime and the part $\langle T_{ik}\rangle _{%
\mathrm{b}}$ is induced by the brane. As the VEV of the field square in free
AdS spacetime does not depend on spacetime point, the corresponding VEV is
found from the formula $\langle 0_{S}|T_{i}^{k}|0_{S}\rangle =\delta
_{i}^{k}m^{2}\langle 0_{S}|\varphi ^{2}|0_{S}\rangle $ by taking into
account formula (\ref{phi2AdS}). For a conformally coupled massless scalar
field and for even values $D$ the renormalized free AdS part in the VEV of
the energy-momentum tensor vanishes. For odd values of $D$, this part is
completely determined by the trace anomaly (see, for instance, \cite{Birrell}%
).

Substituting the expressions of the Wightman function and the VEV of the
field square into formula (\ref{mvevEMT}), for the part of the
energy-momentum tensor induced by the brane one obtains
\begin{equation}
\langle T_{i}^{k}\rangle _{\mathrm{b}}=-\frac{k_{D}^{D+1}z^{D}\delta _{i}^{k}%
}{2^{D-2}\pi ^{(D+1)/2}\Gamma \left( \frac{D-1}{2}\right) }\int_{0}^{\infty
}dkk^{D-2}\int_{k}^{\infty }d\lambda \,\lambda \frac{\tilde{I}_{\nu
}(\lambda z_{a})}{\tilde{K}_{\nu }(\lambda z_{a})}\frac{F^{(i)}\left[
k,K_{\nu }(\lambda z)\right] }{\sqrt{\lambda ^{2}-k^{2}}},
\label{EMT1bounda}
\end{equation}%
where for a given function $g(v)$ we use the notations
\begin{eqnarray}
F^{(0)}[k,g(v)] &=&\left( \frac{1}{2}-2\xi \right) \left[ v^{2}g^{\prime
}{}^{2}(v)+\left( D+\frac{4\xi }{4\xi -1}\right) vg(v)g^{\prime }(v)\right.
\label{F0} \\
&&\left. +(v^{2}+\nu ^{2})g^{2}(v)\right] +(z^{2}k^{2}-v^{2})g^{2}(v)  \notag
\\
F^{(i)}[k,g(v)] &=&F^{(0)}\left[ k,g(v)\right] +\left( v^{2}-\frac{%
Dz^{2}k^{2}}{D-1}\right) g^{2}(v),\quad i=1,\ldots ,D-1,  \label{Fi} \\
F^{(D)}[k,g(v)] &=&-\frac{v^{2}}{2}g^{\prime }{}^{2}(v)+\frac{D}{2}\left(
4\xi -1\right) vg(v)g^{\prime }(v)  \notag \\
&&+\frac{1}{2}\left[ v^{2}+\nu ^{2}+2\xi D(D+1)-D^{2}/2\right] g^{2}(v).
\label{FD}
\end{eqnarray}%
As in the case of the field square, these expression may be further
simplified in the model with Poincare invariance along the directions
parallel to the brane for which $u(y)=v(y)$. In this case the tilted
functions in (\ref{EMT1bounda}) do not depend on $k$ and after the
integration one finds%
\begin{equation}
\langle T_{i}^{k}\rangle _{\mathrm{b}}=-\frac{k_{D}^{D+1}z^{D}\delta _{i}^{k}%
}{(4\pi )^{D/2}\Gamma \left( D/2\right) }\int_{0}^{\infty }dxx^{D-1}\,\frac{%
\tilde{I}_{\nu }(xz_{a})}{\tilde{K}_{\nu }(xz_{a})}F^{(i)}[K_{\nu }(xz)],
\label{Tikc}
\end{equation}%
where for a given function $g(v)$ we have introduced the notations
\begin{eqnarray}
F^{(i)}[g(v)] &=&\left( \frac{1}{2}-2\xi \right) \left[ v^{2}g^{\prime
2}(v)+\left( D+\frac{4\xi }{4\xi -1}\right) vg(v)g^{\prime }(v)\right.
\notag \\
&&+\left. \left( \nu ^{2}+v^{2}+\frac{2v^{2}}{D(4\xi -1)}\right) g^{2}(v)%
\right] ,\quad i=0,1,\ldots ,D-1,  \label{Finew} \\
F^{(D)}[g(v)] &=&-\frac{v^{2}}{2}g^{\prime }{}^{2}(v)+\frac{D}{2}\left( 4\xi
-1\right) vg(v)g^{\prime }(v)  \notag \\
&&+\frac{1}{2}\left[ v^{2}+\nu ^{2}+2\xi D(D+1)-D^{2}/2\right] g^{2}(v).
\label{FDnew}
\end{eqnarray}%
It can be explicitly checked that the VEVs given by (\ref{Tikc}) satisfy the
continuity equation $\nabla _{k}\langle T_{i}^{k}\rangle _{\mathrm{b}}=0$,
which for the geometry under consideration takes the form%
\begin{equation}
z^{D+1}\partial _{z}(z^{-D}\langle T_{D}^{D}\rangle _{\mathrm{b}})+D\langle
T_{0}^{0}\rangle _{\mathrm{b}}=0.  \label{conteq1}
\end{equation}%
For a conformally coupled massless scalar field one has $\nu =1/2$ and from
formulae (\ref{Finew}), (\ref{FDnew}) it follows that $F^{(i)}[K_{\nu
}(x)]=F^{(D)}[K_{\nu }(x)]=0$. Hence, in this case the brane induced parts
in the VEVs of the energy-momentum tensor vanish. Note that for a
conformally coupled scalar and for even values $D$ the conformal anomaly is
absent and the free AdS part in the vacuum energy-momentum tensor vanishes
as well.

For large distances from the brane, $z\gg z_{a}$, introducing a new
integration variable $y=xz$ we expand the integrand over $z_{a}/z$. To the
leading order this leads to the result%
\begin{equation}
\langle T_{i}^{k}\rangle _{\mathrm{b}}=-\frac{k_{D}^{D+1}\delta
_{i}^{k}(z_{a}/z)^{2\nu }}{2^{D+2\nu -1}\pi ^{D/2}\Gamma \left( D/2\right)
\nu \Gamma ^{2}(\nu )}\frac{\mathcal{R}(a)+\nu }{\mathcal{R}(a)-\nu }%
\int_{0}^{\infty }dx\,x^{D+2\nu -1}F^{(i)}[K_{\nu }(x)].  \label{TikLargez}
\end{equation}%
The integrals in this formula may be evaluated by using the formulae from
\cite{Prud86}. Note that the free AdS parts in the VEVs of both field square
and the energy-momentum tensor do not depend on the spacetime point and,
hence, at large distances from the brane they dominate in the total VEVs.
Noting that in the limit of strong gravitational field in the region outside
the brane, corresponding to large values $k_{D}$, one has $%
z/z_{a}=e^{k_{D}(y-a)}\gg 1$, we see that formulae (\ref{phi2cextfar}), (\ref%
{TikLargez}) also describe the asymptotic behavior of the brane induced VEVs
in this limit. Hence, in the limit of strong gravitational field, for the
points not too close to the brane boundary, the brane induced parts are
exponentially suppressed. The free AdS parts behave as $k_{D}^{D-1}$ and
their contribution dominates for strong gravitational fields.

\section{Model with flat spacetime inside the brane}

\label{sec:flowerpot}

As an application of the general results given above let us
consider a simple example assuming that the spacetime inside the
brane is flat. The corresponding models for the cosmic string and
global monopole cores were considered in Refs. \cite{Alle90} and
are known as flower-pot models. As we will see below, this model
is exactly solvable in the sense that the VEVs inside the brane
are expressed in terms of closed analytic formulae. This model is
also of interest from a physical point of view as it reduces to
the standard RS 1-brane model in the limit of the zero thickness
brane and, hence, is the simplest generalization of the latter.
From the continuity condition on the brane boundary it follows
that in coordinates $(x^{\mu },y)$
the interior line element has the form%
\begin{equation}
ds^{2}=e^{-2k_{D}a}\eta _{\mu \nu }dx^{\mu }dx^{\nu }-dy^{2}.
\label{intflow}
\end{equation}%
In the rescaled coordinates%
\begin{equation}
X^{\mu }=e^{-k_{D}a}x^{\mu },  \label{Xmu}
\end{equation}%
this line element is written in the standard Minkowskian form. From the
matching conditions (\ref{matchcond1}), (\ref{matchcond2}) we find the
corresponding surface energy-momentum tensor with the non-zero components%
\begin{equation}
\tau _{i}^{k}=\frac{D-1}{8\pi G}k_{D}\delta _{i}^{k},\;i=0,1,\ldots ,D-1.
\label{surfemtflow}
\end{equation}%
The corresponding surface energy density is positive. Lets us consider the
VEVs in the exterior and interior regions separately.

\subsection{Exterior region}

\label{subsec:flpotExt}

For the simple model under consideration the radial equation (\ref{eqforfn})
is easily solved in the interior region. The corresponding radial
eigenfunction with $Z_{2}$-symmetry, $R(-y,k,\lambda )=R(y,k,\lambda )$, has
the form%
\begin{equation}
R(y,k,\lambda )=C_{\nu }\cos (k_{y}y),\;k_{y}^{2}=\lambda
^{2}e^{2k_{D}a}-m^{2}.  \label{Rlflow}
\end{equation}%
The coefficient $C_{\nu }$ is found from formula (\ref{normcoefRl})%
\begin{equation}
C_{\nu }^{2}=\frac{2z_{a}^{D}k_{D}^{D-1}\lambda }{\pi ^{2}\cos ^{2}(k_{y}a)%
\left[ \bar{J}_{\nu }^{2}(\lambda z_{a})+\bar{Y}_{\nu }^{2}(\lambda z_{a})%
\right] },  \label{Cnu}
\end{equation}%
with the barred notation for the cylindrical functions%
\begin{equation}
\bar{F}(z)\equiv zF^{\prime }(z)+\left[ \frac{D}{2}-2\xi D+\frac{k_{y}}{k_{D}%
}\tan (k_{y}a)\right] F(z).  \label{barredflow}
\end{equation}%
As a result, the parts in the Wightman function, in the VEVs of the field
square and the energy-momentum tensor induced by the brane are given by
formulae (\ref{regWightout}), (\ref{phi2cext}) and (\ref{Tikc})
respectively, where the tilted notations for the modified Bessel functions
are defined by (\ref{Barrednotmod}) with the coefficient%
\begin{equation}
\mathcal{R}(a,k,x)=\frac{D}{2}-2\xi D-\sqrt{x^{2}+m^{2}/k_{D}^{2}}\tanh
(ak_{D}\sqrt{x^{2}+m^{2}/k_{D}^{2}}).  \label{Rcalflowex}
\end{equation}%
By using the recurrence relation for the function $K_{\nu }(x)$, it can be
seen that under the condition%
\begin{equation}
\nu +2\xi D-D/2+(m/k_{D})\tanh (am)\geqslant 0  \label{condstab}
\end{equation}
one has $\tilde{K}_{\nu }(x)<0$ and in the model under consideration there
are no modes with imaginary $\lambda $. In particular, this condition is
satisfied for minimally and conformally coupled fields. For large distances
from the brane the asymptotic behavior of the parts induced by the brane is
given by formulae (\ref{phi2cextfar}) and (\ref{TikLargez}) for the field
square and the energy-momentum tensor respectively. Now in these formulae we
have
\begin{equation}
\mathcal{R}(a)=D/2-2\xi D-(m/k_{D})\tanh (am).  \label{RaFlower}
\end{equation}%
Comparing (\ref{Rcalflowex}) with (\ref{replaceR}), we see that in the limt $%
a\rightarrow 0$ from the results of the model with flat interior spacetime
the corresponding formulae in the RS 1-brane model with a zero thickness
brane are obtained.

As it was mentioned before the VEVs of the field square and the
energy-momentum tensor diverge on the boundary of the brane. To find the
leading terms in the corresponding asymptotic expansions for points near the
boundary, we note that for these points the main contribution into the
integrals come from large values $x$ and we can replace the modified Bessel
functions by corresponding asymptotic formulae for large values of the
argument. In this way, assuming that $y-a\ll a$, one finds%
\begin{equation}
\langle \varphi ^{2}\rangle _{\mathrm{b}}\sim \frac{k_{D}A_{D}}{2^{D+5}\pi
^{(D+1)/2}}\frac{\Gamma ((D-1)/2)}{(D-2)(y-a)^{D-2}},\;D>2,  \label{phi2near}
\end{equation}%
with the notation%
\begin{equation}
A_{D}=4D-D^{2}+1+4\xi D(D-3)-4m^{2}/k_{D}^{2}.  \label{AD}
\end{equation}%
For $D\leqslant 2$ the VEV of the field square is finite on the core
boundary. Note that in the model with zero thickness brane located at $y=0$,
the corresponding VEVs near the brane behave as $y^{1-D}$.

The asymptotic behavior for the energy-momentum tensor is found in the
similar way. To the leading order we have (no summation over $i$)%
\begin{equation}
\langle T_{i}^{i}\rangle _{\mathrm{b}}\sim -\frac{k_{D}A_{D}(\xi -\xi _{D})}{%
2^{D+4}\pi ^{(D+1)/2}}\frac{\Gamma ((D+1)/2)}{(y-a)^{D}},\;i=0,1,\ldots ,D-1.
\label{T00near}
\end{equation}%
The behavior for the radial stress is found most easily from the continuity
equation (\ref{conteq1}). This leads to the result%
\begin{equation}
\langle T_{D}^{D}\rangle _{\mathrm{b}}\sim \frac{Dk_{D}(y-a)}{D-1}\langle
T_{0}^{0}\rangle _{\mathrm{b}}.  \label{TDDnear}
\end{equation}%
For a conformally coupled field the most easy way to find the asymptotic
behavior is to note that one has the following relation for the trace of the
energy-momentum tensor%
\begin{equation*}
\langle T_{i}^{i}\rangle _{\mathrm{b}}=D(\xi -\xi _{D})\nabla _{l}\nabla
^{l}\langle \varphi ^{2}\rangle _{\mathrm{b}}+m^{2}\langle \varphi
^{2}\rangle _{\mathrm{b}}.
\end{equation*}%
From here and the continuity equation it follows that for a conformally
coupled scalar field we have (no summation over $i$)%
\begin{equation}
\langle T_{i}^{i}\rangle _{\mathrm{b}}\sim \frac{D-3}{Dk_{D}(y-a)}\langle
T_{D}^{D}\rangle _{\mathrm{b}}\sim (m^{2}/D)\langle \varphi ^{2}\rangle _{%
\mathrm{b}},\;i=0,1,\ldots ,D-1,\;D>3.  \label{Tiinearconf}
\end{equation}%
For $D=3$ the radial stress $\langle T_{D}^{D}\rangle _{\mathrm{b}}$
diverges logarithmically. In the case $D=2$ the corresponding VEVs are
finite on the boundary of the brane.

Now we turn to the investigation of the general formulae for the
brane-induced VEVs in the limit $m/k_{D}\gg 1$. In this limit one has $\nu
\approx m/k_{D}$ and the order of the modified Bessel functions in formulae (%
\ref{phi2cext}) and (\ref{Tikc}) is large. Introducing a new integration
variable $x\rightarrow \nu x$, we replace the modified Bessel functions by
the leading terms of the corresponding uniform asymptotic expansions for
large values of the order (see, for instance, \cite{abramowiz}). By this way
it can be seen that the main contribution into the integral comes from the
lower limit and to the leading order we find%
\begin{equation}
\langle \varphi ^{2}\rangle _{\mathrm{b}}\approx \frac{%
k_{D}^{D-1}(z/z_{a})^{D}(m/k_{D})^{D/2-1}}{2^{D/2+2}\pi
^{D/2}(z^{2}/z_{a}^{2}-1)^{D/2}}\frac{1-\tanh (am)}{1+\tanh (am)}\exp \left[
-2(m/k_{D})\ln (z/z_{a})\right] .  \label{phi2blargem}
\end{equation}%
As we could expect, in this limit the VEV is exponentially suppressed. By
the similar way, for the corresponding VEV of the energy-momentum tensor we
obtain (no summation $i$)%
\begin{equation}
\langle T_{i}^{i}\rangle _{\mathrm{b}}\approx (1-4\xi )m^{2}\langle \varphi
^{2}\rangle _{\mathrm{b}},\;\langle T_{D}^{D}\rangle _{\mathrm{b}}\approx
\frac{Dk_{D}}{2m}\langle T_{0}^{0}\rangle _{\mathrm{b}}.  \label{Tiklargem}
\end{equation}%
The last relation is found most easily from equation (\ref{conteq1}).

\begin{figure}[tbph]
\begin{center}
\epsfig{figure=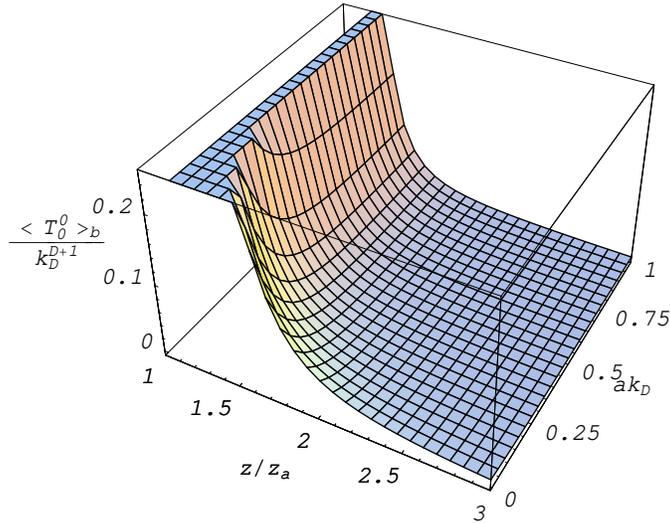,width=9.cm,height=8.5cm}
\end{center}
\caption{The part in the VEV of the energy density, $k_{D}^{-D-1}\langle
T_{0}^{0}\rangle _{\mathrm{b}}$, induced by the brane as a function on $%
z/z_{a}$ and $ak_{D}$ for a minimally coupled massless scalar field in $D=4$%
. }
\label{figT00ex}
\end{figure}

\begin{figure}[tbph]
\begin{center}
\epsfig{figure=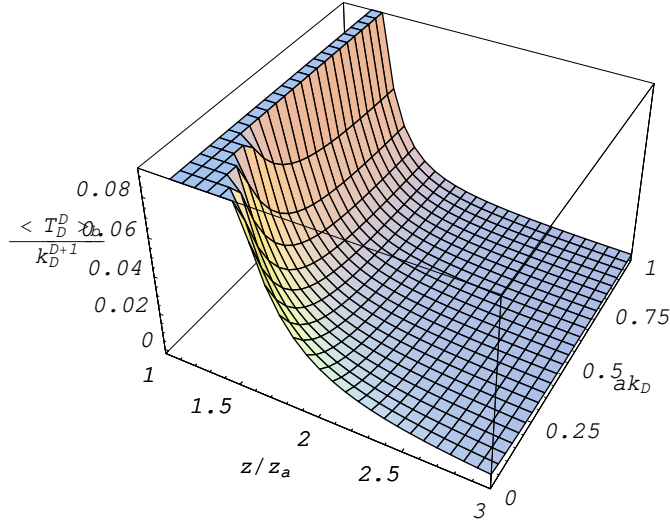,width=9.cm,height=8.5cm}
\end{center}
\caption{The part in the VEV of the radial stress, $k_{D}^{-D-1}\langle
T_{D}^{D}\rangle _{\mathrm{b}}$, induced by the brane as a function on $%
z/z_{a}$ and $ak_{D}$ for a minimally coupled massless scalar field in $D=4$%
. }
\label{figTDDex}
\end{figure}

In figures \ref{figT00ex} and \ref{figTDDex} we have plotted the dependence
of the brane induced parts in the VEVs of the energy density and radial
stress on $z/z_{a}$ and $ak_{D}$ for a minimally coupled massless scalar
field ($\xi =0$) in the case $D=4$. The first of these parameters is related
to the distance from the boundary of the brane by the formula $z/z_{a}=\exp
[k_{D}(y-a)]$. The curves for $a=0$ correspond to the VEVs in the RS 1-brane
model. Recall that for a conformally coupled massless scalar field the brane
induced VEVs vanish. Note that in $D=4$ the conformal anomaly is absent and
for massless scalar fields the VEV of the energy-momentum tensor in the free
AdS spacetime is zero.

\subsection{Interior region}

\label{subsec:flpotinter}

Now let us consider the vacuum polarization effects inside the brane for the
model with flat interior. The corresponding eigenfunctions have the form
given by Eq. (\ref{eigfunc1}) with $f_{\lambda }(y)=R(y,k,\lambda )$ and the
function $R(y,k,\lambda )$ is defined by formula (\ref{Rlflow}).
Substituting the eigenfunctions into the mode sum formula for the
corresponding Wightman function one finds%
\begin{eqnarray}
\langle 0|\varphi (x)\varphi (x^{\prime })|0\rangle &=&\frac{%
z_{a}^{D}k_{D}^{D-1}}{\pi ^{2}}\int d\mathbf{k\,}\frac{e^{i\mathbf{k}\cdot (%
\mathbf{x}-\mathbf{x}^{\prime })}}{(2\pi )^{D-1}}  \notag \\
&&\times \int_{0}^{\infty }d\lambda \,\frac{\lambda \cos (k_{y}y)\cos
(k_{y}y^{\prime })e^{-i\sqrt{k^{2}+\lambda ^{2}}(t-t^{\prime })}}{\cos
^{2}(k_{y}a)\sqrt{k^{2}+\lambda ^{2}}\left[ \bar{J}_{\nu }^{2}(\lambda
z_{a})+\bar{Y}_{\nu }^{2}(\lambda z_{a})\right] }.  \label{WFflowin}
\end{eqnarray}%
For the evaluation of the expression on the right we use the relation%
\begin{equation}
\frac{\cos ^{-2}(k_{y}a)}{\bar{J}_{\nu }^{2}(\lambda z_{a})+\bar{Y}_{\nu
}^{2}(\lambda z_{a})}=\frac{\pi k_{D}}{2k_{y}}\left[ 1-\frac{1}{2}%
\sum_{s=1,2}\frac{C\{e^{i\delta _{s}k_{y}a},H_{\nu }^{(s)}(\lambda z_{a})\}}{%
C\{\cos (k_{y}a),H_{\nu }^{(s)}(\lambda z_{a})\}}\right] .  \label{Flident3}
\end{equation}%
where $\delta _{s}=(-1)^{s+1}$, and for given functions $f(z)$ and $g(z)$ we
have introduced the notation%
\begin{equation}
C\left\{ f(u),g(v)\right\} =vf(u)g^{\prime }(v)+\left[ \left( \frac{D}{2}%
-2\xi D\right) f(u)-\frac{u}{ak_{D}}f^{\prime }(u)\right] g(v),  \label{Cfg}
\end{equation}%
with $u=k_{y}a$ and $v=\lambda z_{a}$.

\begin{figure}[tbph]
\begin{center}
\epsfig{figure=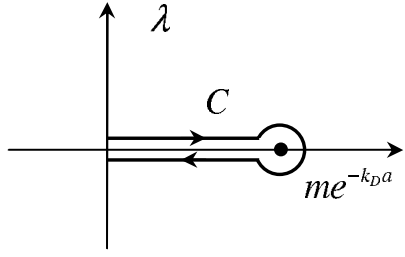,width=5.5cm,height=3.5cm}
\end{center}
\caption{Integration contour $C$ in the complex $\protect\lambda $-plane.}
\label{figCont}
\end{figure}

By taking into account identity (\ref{Flident3}), the Wightman function is
written in the form%
\begin{eqnarray}
\langle 0|\varphi (x)\varphi (x^{\prime })|0\rangle &=&(z_{a}k_{D})^{D}\int d%
\mathbf{k\,}\frac{e^{i\mathbf{k}\cdot (\mathbf{x}-\mathbf{x}^{\prime })}}{%
(2\pi )^{D}}\int_{0}^{\infty }d\lambda \,\frac{\lambda e^{-i\sqrt{%
k^{2}+\lambda ^{2}}(t-t^{\prime })}}{k_{y}\sqrt{k^{2}+\lambda ^{2}}}  \notag
\\
&&\times \cos (k_{y}y)\cos (k_{y}y^{\prime })\left[ 1-\frac{1}{2}\sum_{s=1,2}%
\frac{C\{e^{i\delta _{s}k_{y}a},H_{\nu }^{(s)}(\lambda z_{a})\}}{C\{\cos
(k_{y}a),H_{\nu }^{(s)}(\lambda z_{a})\}}\right] .  \label{WFfl2}
\end{eqnarray}%
To transform the $\lambda $-integral with the first term in the square
brackets, we present this integral as a sum of the integrals over the
intervals $(0,me^{-k_{D}a})$ and $(me^{-k_{D}a},\infty )$. In the second
integral changing the integration variable to $k_{y}$ we find%
\begin{eqnarray}
&&\int_{0}^{\infty }d\lambda \,\frac{\lambda e^{-i\sqrt{k^{2}+\lambda ^{2}}%
(t-t^{\prime })}}{k_{y}\sqrt{k^{2}+\lambda ^{2}}}\cos (k_{y}y)\cos
(k_{y}y^{\prime })=e^{-2k_{D}a}\int_{0}^{\infty }dk_{y}\,\frac{e^{-i\sqrt{%
k^{2}+\lambda ^{2}}(t-t^{\prime })}}{\sqrt{k^{2}+\lambda ^{2}}}  \notag \\
&&\quad \times \cos (k_{y}y)\cos (k_{y}y^{\prime
})-i\int_{0}^{me^{-k_{D}a}}d\lambda \,\frac{\lambda e^{-i\sqrt{k^{2}+\lambda
^{2}}(t-t^{\prime })}}{\eta (\lambda )\sqrt{k^{2}+\lambda ^{2}}}\cosh (y\eta
(\lambda ))\cosh (y^{\prime }\eta (\lambda )),  \label{Iterm}
\end{eqnarray}%
with the notation $\eta (\lambda )=\sqrt{m^{2}-\lambda ^{2}e^{2k_{D}a}}$. In
the first integral on the right of (\ref{Iterm}), $\lambda $ is a function
of $k_{y}$ given by formula (\ref{Rlflow}). Let us consider the $\lambda $%
-integral with the second term in the square brackets of (\ref{WFfl2}). Note
that in accordance with formula (\ref{Rlflow}) for $k_{y}$, the
corresponding integrand has branch point at $\lambda =me^{-k_{D}a}$. For
definiteness we will assume that this point is circled from above by a
semicircle of small radius in the complex $\lambda $-plane. With this
choice, assuming that $|y|+|y^{\prime }|+|t-t^{\prime }|<2a$, we rotate the
integration contour in the complex plane $\lambda $ by the angle $\pi /2$
for $s=1$ and by the angle $-\pi /2$ for $s=2$. In the second case the
integral with $s=2$ is equal to the corresponding integral over the negative
part of the imaginary axis plus the integral over the contour $C$ depicted
in figure \ref{figCont}. By taking into account that on the upper/lower
parts of the contour $C$ we have $k_{y}=\pm i\eta (\lambda )$, for this
contribution one finds%
\begin{eqnarray}
&&-\int_{C}d\lambda \,\frac{\lambda \cos (k_{y}y)\cos (k_{y}y^{\prime })e^{-i%
\sqrt{k^{2}+\lambda ^{2}}(t-t^{\prime })}}{2k_{y}\sqrt{k^{2}+\lambda ^{2}}}%
\frac{C\{e^{-ik_{y}a},H_{\nu }^{(2)}(\lambda z_{a})\}}{C\{\cos
(k_{y}a),H_{\nu }^{(2)}(\lambda z_{a})\}}  \notag \\
&&\quad =i\int_{0}^{me^{-k_{D}a}}d\lambda \,\frac{\lambda e^{-i\sqrt{%
k^{2}+\lambda ^{2}}(t-t^{\prime })}}{\eta (\lambda )\sqrt{k^{2}+\lambda ^{2}}%
}\cosh (y\eta (\lambda ))\cosh (y^{\prime }\eta (\lambda )).
\label{hashCcont}
\end{eqnarray}%
Now we see that the the integral over the contour $C$ cancels the second
term on the right of (\ref{Iterm}). In the integrals over the imaginary axis
the parts over the intervals $(0,ik)$ and $(0,-ik)$ cancel out. After
introducing the modified Bessel functions, one can see that the Wightman
function is presented in the form
\begin{equation}
\langle 0|\varphi (x)\varphi (x^{\prime })|0\rangle =G_{0}(x,x^{\prime
})+G_{1}(x,x^{\prime }),  \label{G01}
\end{equation}%
where%
\begin{equation}
G_{0}(x,x^{\prime })=(z_{a}k_{D})^{D-2}\int \frac{d\mathbf{k}}{(2\pi )^{D}}%
e^{i\mathbf{k}\cdot (\mathbf{x}-\mathbf{x}^{\prime })}\int_{0}^{\infty
}dk_{y}\,\frac{e^{-i\sqrt{k^{2}+\lambda ^{2}}(t-t^{\prime })}}{\sqrt{%
k^{2}+\lambda ^{2}}}\cos (k_{y}y)\cos (k_{y}y^{\prime }),  \label{G0}
\end{equation}%
and%
\begin{eqnarray}
G_{1}(x,x^{\prime }) &=&-\frac{(z_{a}k_{D})^{D}}{(2\pi )^{D}}\int d\mathbf{%
k\,}e^{i\mathbf{k}\cdot (\mathbf{x}-\mathbf{x}^{\prime })}\int_{k}^{\infty
}dx\,\frac{xC\{e^{-\varkappa (x)a},K_{\nu }(xz_{a})\}}{C\{\cosh (\varkappa
(x)a),K_{\nu }(xz_{a})\}}  \notag \\
&&\times \frac{\cosh (\varkappa (x)y)\cosh (\varkappa (x)y^{\prime })}{%
\varkappa (x)\sqrt{x^{2}-k^{2}}}\cosh [\sqrt{x^{2}-k^{2}}(t-t^{\prime })].
\label{G1}
\end{eqnarray}%
In (\ref{G1}), $\varkappa (x)=\sqrt{x^{2}e^{2k_{D}a}+m^{2}}$, and we have
used the notation (\ref{Cfg}) with $g(v)=K_{\nu }(v)$ and $f(u)=e^{-u},\cosh
u$ for the numerator and denominator, respectively.

In (\ref{G0}) introducing a new integration variable $\mathbf{K}=\mathbf{k}%
e^{k_{D}a}$ we see that $G_{0}(x,x^{\prime })$ is the Wightman function in
the Minkowski spacetime in coordinates (\ref{intflow}) orbifolded along the $%
y$-direction:%
\begin{equation}
G_{0}(x,x^{\prime })=\int d\mathbf{K\,}\frac{e^{i\mathbf{K}\cdot (\mathbf{X}-%
\mathbf{X}^{\prime })}}{(2\pi )^{D}}\int_{0}^{\infty }dk_{y}\,\frac{e^{-i%
\sqrt{K^{2}+k_{y}^{2}+m^{2}}(X^{0}-X^{0\prime })}}{\sqrt{%
K^{2}+k_{y}^{2}+m^{2}}}\cos (k_{y}y)\cos (k_{y}y^{\prime }),  \label{G02}
\end{equation}%
where $\mathbf{X}=(X^{1},X^{2},\ldots ,X^{D-1})$ with coordinates $X^{\mu }$
defined by relation (\ref{Xmu}). This function differs by the factor 1/2
from the Wightman function for a plate in the Minkowski spacetime located at
$y=0$ on which the field obeys the Neumann boundary condition. Evaluating
the integrals, function (\ref{G02}) is presented in the form%
\begin{eqnarray}
G_{0}(x,x^{\prime }) &=&\frac{1}{2}G_{M}(x,x^{\prime })+\frac{m^{(D-1)/2}}{%
2(2\pi )^{(D+1)/2}}  \notag \\
&&\times \frac{K_{(D-1)/2}\left( m\sqrt{|\mathbf{X}-\mathbf{X}^{\prime
}|^{2}+(y+y^{\prime })^{2}-(X^{0}-X^{0\prime })^{2}}\right) }{\left( |%
\mathbf{X}-\mathbf{X}^{\prime }|^{2}+(y+y^{\prime })^{2}-(X^{0}-X^{0\prime
})^{2}\right) ^{(D-1)/4}},  \label{G03}
\end{eqnarray}%
where $G_{M}(x,x^{\prime })$ is the Wightman function for the Minkowski
spacetime. Note that the expression for $G_{M}(x,x^{\prime })/2$ is given by
the same formula as the second term on the right of (\ref{G03}) with the
replacement $y+y^{\prime }\rightarrow y-y^{\prime }$. The term $%
G_{1}(x,x^{\prime })$ on the right of formula (\ref{G01}) is induced by the
AdS geometry in the region $y>a$. For a conformally coupled massless scalar
field one has $\nu =1/2$ and by using definition (\ref{Cfg}) it can be
explicitly checked that $C\{e^{-u},K_{\nu }(u/ak_{D})\}=0$. Hence, in this
case the part $G_{1}(x,x^{\prime })$ of the Wightman function vanishes.

Now we turn to the evaluation of the renormalized VEV for the field square.
The renormalization corresponds to the omission of the part coming from the
Minkowskian Wightman function in (\ref{G03}). As a result the VEV for the
field square is presented in the form%
\begin{equation}
\langle \varphi ^{2}\rangle _{\mathrm{ren}}^{\mathrm{(int)}}=\langle \varphi
^{2}\rangle _{0,\mathrm{ren}}^{\mathrm{(int)}}+\langle \varphi ^{2}\rangle _{%
\mathrm{b}}^{\mathrm{(int)}}.  \label{phi2int}
\end{equation}%
Here the part $\langle \varphi ^{2}\rangle _{0,\mathrm{ren}}^{\mathrm{(int)}}
$ is obtained from the second term on the right of (\ref{G03}) in the
coincidence limit and is given by the formula%
\begin{equation}
\langle \varphi ^{2}\rangle _{0,\mathrm{ren}}^{\mathrm{(int)}}=\frac{m^{D-1}%
}{2(2\pi )^{(D+1)/2}}\frac{K_{(D-1)/2}\left( 2my\right) }{\left( 2my\right)
^{(D-1)/2}}.  \label{phi2int0}
\end{equation}%
The second term in the right hand-side of formula (\ref{phi2int}) is
obtained from (\ref{G1}) taking the coincidence limit of the arguments and
using the formula%
\begin{equation}
\int d\mathbf{k\,}\int_{k}^{\infty }dx\frac{k^{s}g(x)}{\sqrt{x^{2}-k^{2}}}=%
\frac{\pi ^{(D-1)/2}}{\Gamma ((D-1)/2)}B\left( \frac{D-1+s}{2},\frac{1}{2}%
\right) \int_{0}^{\infty }dx\,x^{D+s-2}g(x),  \label{intform}
\end{equation}%
with $B(x,y)$ being the Euler beta function. This gives the result%
\begin{equation}
\langle \varphi ^{2}\rangle _{\mathrm{b}}^{\mathrm{(int)}}=-\frac{(4\pi
)^{-D/2}}{\Gamma (D/2)}\int_{m}^{\infty }dx\,(x^{2}-m^{2})^{D/2-1}\cosh
^{2}(xy)U_{\nu }(x),  \label{phi2intb}
\end{equation}%
with the notation%
\begin{equation}
U_{\nu }(x)=\frac{C\{e^{-ax},K_{\nu }(\sqrt{x^{2}-m^{2}}/k_{D})\}}{C\{\cosh
(ax),K_{\nu }(\sqrt{x^{2}-m^{2}}/k_{D})\}}.  \label{Unu}
\end{equation}%
This part in the VEV of the field square is induced by the AdS geometry in
the exterior region. Note that if condition (\ref{condstab}) is satisfied
the denominator in (\ref{Unu}) is negative. Recall that under this condition
there are no modes with imaginary $\lambda $.

The integral on the right of formula (\ref{phi2intb}) is finite for $|y|<a$
and diverges on the boundary of the brane $|y|=a$. In order to find the
corresponding asymptotic behavior, we note that for points near the boundary
the main contribution comes from large values $x$. By using the
corresponding asymptotic formula for the function $K_{\nu }(x)$, to the
leading order near the boundary $y=a$ we find

\begin{equation}
\langle \varphi ^{2}\rangle _{\mathrm{b}}^{\mathrm{(int)}}\sim -\frac{%
k_{D}(\xi -\xi _{D})}{(4\pi )^{(D+1)/2}}\frac{D\Gamma ((D-1)/2)}{%
(D-2)(a-y)^{D-2}}.  \label{phi2intnear}
\end{equation}%
In the limit $am\gg 1$ the main contribution into the integral in formula (%
\ref{phi2intb}) comes from the lower limit and one finds%
\begin{equation}
\langle \varphi ^{2}\rangle _{\mathrm{b}}^{\mathrm{(int)}}\approx -\frac{%
Bm^{D-1}\cosh ^{2}(my)}{(4\pi am)^{D/2}}e^{-2am},\;B\equiv \frac{D/2-2\xi
D+m/k_{D}-\nu }{D/2-2\xi D-m/k_{D}-\nu }.  \label{phi2amlarge}
\end{equation}%
As we could expect, in this limit the VEVs are exponentially suppressed. For
large values of the AdS curvature assuming that $a,m\gg 1/k_{D}$, from
definition (\ref{Cfg}) it is easily seen that under the conditions $\xi \gg
1/(ak_{D})$, $\xi -\xi _{D}\gg 1/(ak_{D})$ the asymptotic behavior of $%
\langle \varphi ^{2}\rangle _{\mathrm{b}}^{\mathrm{(int)}}$ is obtained
substituting in formula (\ref{phi2intb})%
\begin{equation}
U_{\nu }(x)\approx \frac{2}{e^{2ax}+1}.  \label{Unu1}
\end{equation}%
The corresponding expression coincides with the VEV induced by Dirichlet
boundary located at $y=a$ in the Minkowski spacetime orbifolded along the $y$%
-direction with the fixed point $y=0$. For special cases of minimally and
conformally coupled scalar fields, in the limit under consideration the
leading terms are obtained substituting in (\ref{phi2intb})%
\begin{eqnarray}
U_{\nu }(x) &\approx &-\frac{2}{e^{2ax}-1},\;\xi =0,  \label{Unu2} \\
U_{\nu }(x) &\approx &-\frac{2}{\frac{x+\sqrt{x^{2}-m^{2}}}{x-\sqrt{%
x^{2}-m^{2}}}e^{2ax}-1},\;\xi =\xi _{D}.  \label{Unu3}
\end{eqnarray}%
In the minimally coupled case the corresponding limiting value coincides
with the VEV induced by Neumann boundary located at $y=a$ in the Minkowski
spacetime orbifolded along the $y$-direction with the fixed point $y=0$.

Now let us consider the VEV for the energy-momentum tensor. As in the case
of the field square, it is presented in the form%
\begin{equation}
\langle T_{i}^{k}\rangle _{\mathrm{ren}}^{\mathrm{(int)}}=\langle
T_{i}^{k}\rangle _{0,\mathrm{ren}}^{\mathrm{(int)}}+\langle T_{i}^{k}\rangle
_{\mathrm{b}}^{\mathrm{(int)}},  \label{Tikint}
\end{equation}%
where $\langle T_{ik}\rangle _{0,\mathrm{ren}}^{\mathrm{(int)}}$ is the
vacuum energy-momentum tensor in the Minkowski spacetime in coordinates (\ref%
{intflow}) orbifolded along the $y$-direction and the presence of the part $%
\langle T_{ik}\rangle _{\mathrm{b}}^{\mathrm{(int)}}$ is related to that the
geometry in the region $y>a$ is AdS. For the first part one has (no
summation over $i$)
\begin{equation}
\langle T_{i}^{i}\rangle _{0,\mathrm{ren}}^{\mathrm{(int)}}=\frac{m^{D+1}}{%
(4\pi my)^{(D+1)/2}}\left\{ K_{(D+1)/2}\left( 2my\right) \left( 2\xi
-1\right) +\left( 1-4\xi \right) myK_{(D+3)/2}\left( 2my\right) \right\} ,
\label{Tiiint0n}
\end{equation}%
with $\;i=0,1,\ldots ,D-1$ and $\langle T_{D}^{D}\rangle _{0,\mathrm{ren}}^{%
\mathrm{(int)}}=0$. For a massless scalar field this leads to the result (no
summation over $i$)%
\begin{equation}
\langle T_{i}^{i}\rangle _{0,\mathrm{ren}}^{\mathrm{(int)}}=-\frac{D\left(
\xi -\xi _{D}\right) }{(4\pi )^{(D+1)/2}y^{D+1}}\Gamma \left( \frac{D+1}{2}%
\right) .  \label{Tii0}
\end{equation}%
For the second term on the right of (\ref{Tikint}) we find (no summation
over $i$)%
\begin{eqnarray}
\langle T_{i}^{i}\rangle _{\mathrm{b}}^{\mathrm{(int)}} &=&\frac{(4\pi
)^{-D/2}}{\Gamma (D/2)}\int_{m}^{\infty }dx(x^{2}-m^{2})^{D/2}U_{\nu }(x)
\notag \\
&&\times \left\{ \frac{1}{D}\cosh ^{2}(xy)+\frac{\left( 4\xi -1\right) x^{2}%
}{x^{2}-m^{2}}\left[ \cosh ^{2}(xy)-\frac{1}{2}\right] \right\} ,
\label{T00intb} \\
\langle T_{D}^{D}\rangle _{\mathrm{b}}^{\mathrm{(int)}} &=&-\frac{(4\pi
)^{-D/2}}{2\Gamma (D/2)}\int_{m}^{\infty
}dx\,(x^{2}-m^{2})^{D/2-1}x^{2}U_{\nu }(x),  \label{TDDintb}
\end{eqnarray}%
with $i=0,1,\ldots ,D-1$, and the function $U_{\nu }(x)$ is defined by
formula (\ref{Unu}). Note that the radial stress inside the brane does not
depend on spacetime point. This result could be also obtained directly from
the continuity equation. As it has been mentioned before, for a conformally
coupled massless scalar we have $U_{\nu }(x)=0$ and, hence, the parts in the
VEV of the energy-momentum tensor given by formulae (\ref{T00intb}),(\ref%
{TDDintb}) vanish. From formula (\ref{Tii0}) it follows that in this case
the part $\langle T_{i}^{i}\rangle _{0,\mathrm{ren}}^{\mathrm{(int)}}$
vanishes as well.

Now let us consider the VEV of the energy-momentum tensor near the brane
core. In the way similar to that for the exterior region we find (no
summation over $i$)%
\begin{equation}
\langle T_{i}^{i}\rangle _{\mathrm{b}}^{\mathrm{(int)}}\sim \frac{D-1}{%
Dk_{D}(y-a)}\langle T_{D}^{D}\rangle _{\mathrm{b}}^{\mathrm{(int)}}\sim
\frac{Dk_{D}(\xi -\xi _{D})^{2}}{2^{D}\pi ^{(D+1)/2}}\frac{\Gamma ((D+1)/2)}{%
(a-y)^{D}},  \label{Tiiintnear}
\end{equation}%
with $i=0,1,\ldots ,D-1$. For a conformally coupled scalar field the
corresponding asymptotic behavior is given by formulae (\ref{Tiinearconf}).
In the limit $am\gg 1$ to the leading order one has (no summation over $i$)%
\begin{eqnarray}
\langle T_{i}^{i}\rangle _{\mathrm{b}}^{\mathrm{(int)}} &\approx &\frac{%
Bm^{D+1}e^{-2am}}{2(4\pi am)^{D/2}}(4\xi -1)\left[ 2\cosh ^{2}(my)-1\right] ,
\label{Tiilargeam} \\
\langle T_{D}^{D}\rangle _{\mathrm{b}}^{\mathrm{(int)}} &\approx &-\frac{%
Bm^{D+1}e^{-2am}}{2(4\pi am)^{D/2}}.  \label{TDDlargeam}
\end{eqnarray}%
For large values of the AdS curvature, $a,m\gg 1/k_{D}$, the leading terms
for the VEV of the energy-momentum tensor are obtained from formulae (\ref%
{T00intb}),(\ref{TDDintb}), by making substitutions (\ref{Unu1}) for $\xi
\gg 1/(ak_{D})$, $\xi -\xi _{D}\gg 1/(ak_{D})$, and (\ref{Unu2}),(\ref{Unu3}%
) for minimally and conformally coupled scalars.

\begin{figure}[tbph]
\begin{center}
\epsfig{figure=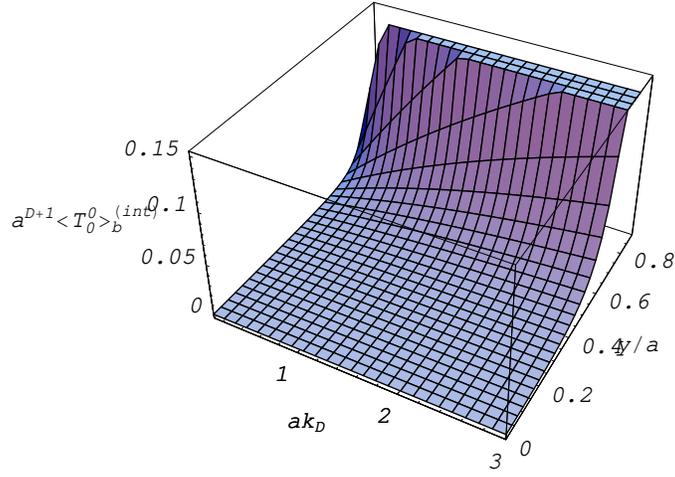,width=9.cm,height=8.5cm}
\end{center}
\caption{The energy density inside the brane, $a^{D+1}\langle
T_{0}^{0}\rangle _{\mathrm{b}}^{\mathrm{(int)}}$, induced by the AdS
geometry in the exterior region as a function on $ak_{D}$ and $y/a$ for a
minimally coupled massless scalar field in $D=4$. }
\label{fig4}
\end{figure}

In figure \ref{fig4} we have plotted the dependence of the part in the VEV
of the energy density induced by the exterior AdS geometry in the region
inside the brane as a function of $y/a$ and $ak_{D}$ for a minimally coupled
massless scalar field in the case $D=4$.

\begin{figure}[tbph]
\begin{center}
\epsfig{figure=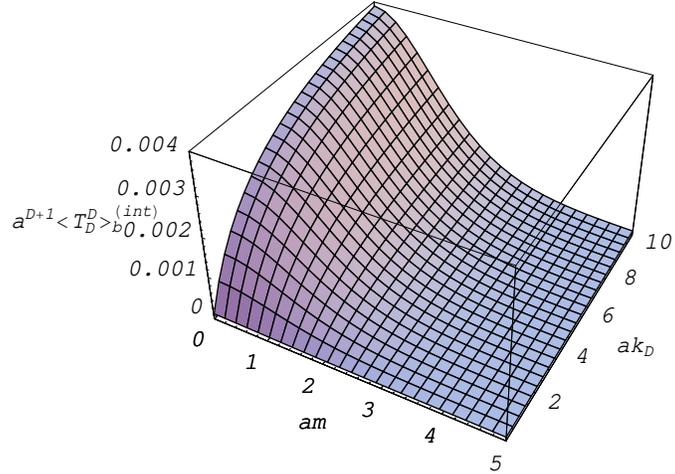,width=9.cm,height=8.5cm}
\end{center}
\caption{The part in the radial stress inside the brane, $a^{D+1}\langle
T_{D}^{D}\rangle _{\mathrm{b}}^{\mathrm{(int)}}$, induced by the AdS
geometry of the exterior region as a function on $am$ and $ak_{D}$ for a
minimally coupled scalar field in $D=4$. }
\label{fig5}
\end{figure}

\begin{figure}[tbph]
\begin{center}
\epsfig{figure=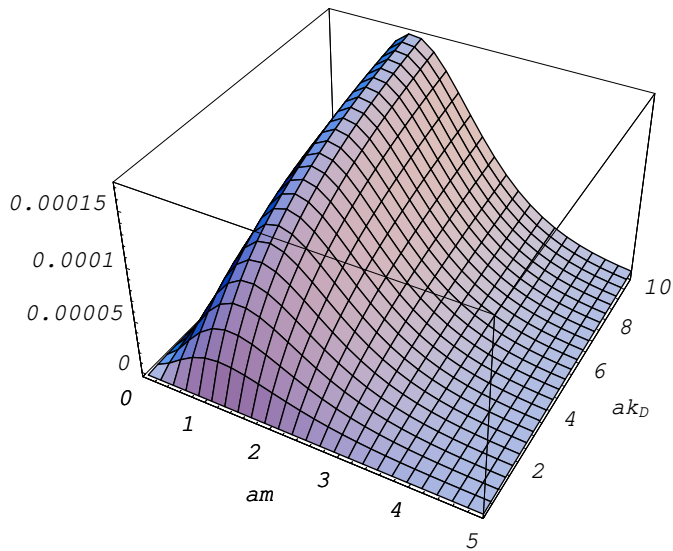,width=9.cm,height=8.5cm}
\end{center}
\caption{The same as in figure \protect\ref{fig5} for a conformally coupled
scalar field. }
\label{fig6}
\end{figure}

The corresponding radial stress does not depend on $y$ and is depicted in
figures \ref{fig5}, \ref{fig6} as a function of the parameters $am$ and $%
ak_{D}$ for minimally and conformally coupled fields, respectively. We
recall that for a conformally coupled massless scalar field the
corresponding VEVs vanish for both field square and the energy-momentum
tensor. The perpendicular interior vacuum force acting per unit surface of
the brane boundary is determined by $-\langle T_{D}^{D}\rangle _{\mathrm{b}%
}^{\mathrm{(int)}}$. As it is seen from figures \ref{fig5}, \ref{fig6} for
minimally and conformally coupled scalars these forces tend to decrease the
brane thickness.

\section{Conclusion}

\label{sec:conc}

In braneworld models the investigation of quantum effects is of considerable
phenomenological interest, both in particle physics and in cosmology. In the
present paper we have considered the one-loop vacuum effects for a massive
scalar field with general curvature coupling parameter induced by a $Z_{2}$%
-symmetric thick brane on the $(D+1)$-dimensional AdS bulk. The previous
papers on the investigation of the vacuum polarization by the gravitational
field of the brane are mainly concerned with the idealized thin brane model,
where the curvature has singularity at the location of the brane. Here we
consider the general plane symmetric static model of the brane with finite
thickness, described by the line element (\ref{metricinside}). Among the
most important characteristics of the vacuum, which carry information about
the internal structure of the brane, are the VEVs for the field square and
the energy-momentum tensor. In order to obtain these expectation values we
first construct the positive frequency Wightman function. In the region
outside the brane this function is presented as a sum of two distinct
contributions. The first one corresponds to the Wightman function in the
free AdS geometry and the second one is induced by the brane. The latter is
given by formula (\ref{regWightout}), where the tilted notation is defined
by formula (\ref{Barrednotmod}) with the coefficient from (\ref{Rlcal}) for
the model without an infinitely thin shell on the brane boundary. This
coefficient is determined by the radial part of the interior eigenfunctions
and describes the influence of the core properties on the vacuum
characteristics in the exterior region. In the case of the model with a thin
shell on the boundary of the brane, the derivatives of the metric tensor
components are discontinuous on the brane surface. This leads to the delta
function type contribution in the Ricci scalar and, hence in the equation
for the radial eigenfunctions in the case of the non-minimally coupled
scalar field. As a result, the radial eigenfunctions have a discontinuity in
their slope at the brane boundary. This leads to an additional term in the
coefficient of the tilted notation which is proportional to the trace of the
surface energy-momentum tensor (see Eq. (\ref{newRlcal})).

By using the formula for the Wightman function, in section \ref{sec:Outside}
we have investigated the influence of the non-trivial internal structure of
the brane on the VEVs of the field square and the energy-momentum tensor.
The parts in these VEVs induced by the brane are directly obtained from the
corresponding part of the Wightman function for the case of the field square
and by applying on this function a certain second-order differential
operator and taking the coincidence limit for the energy-momentum tensor. In
the general plane symmetric model for the brane interior these parts are
given by formulae (\ref{phi2cext0}) and (\ref{EMT1bounda}) for the field
square and the energy-momentum tensor respectively. These formulae are
further simplified for models with Poincare invariance along the directions
parallel to the brane taking the form (\ref{phi2cext}), (\ref{Tikc}). For a
conformally coupled massless scalar field the corresponding energy-momentum
tensor vanishes. The parts in the VEVs of the field square and
energy-momentum tensor induced by the brane diverge on the boundary of the
brane. The surface divergences in the VEVs of the local observables are
well--known in quantum field theory with boundaries and are investigated for
various boundary geometries. At large distances from the brane the brane
induced VEVs are suppressed by the factor $e^{-2\nu k_{D}y}$. In the limit
of strong gravitational fields corresponding to large values of the AdS
energy scale $k_{D}$, for points not too close to the brane the parts in the
VEVs induced by the brane behave as $k_{D}^{D\pm 1}e^{-2\nu k_{D}(y-a)}$
with upper/lower sign corresponding to the energy-momentum tensor/field
square. In this case the relative contribution of the brane induced effects
are exponentially suppressed with respect to the free AdS part.

As an application of the general results, in section \ref{sec:flowerpot} we
have considered a simple model with flat spacetime in the region inside the
brane. The corresponding surface energy-momentum tensor on the boundary of
the brane is obtained from the matching conditions and has the form given by
Eq. (\ref{surfemtflow}). The brane induced parts of the exterior VEVs in
this model are obtained from the general results by taking the function in
the coefficient of the tilted notation from Eq. (\ref{Rcalflowex}). We have
also investigated the vacuum densities inside the brane. Though the
spacetime geometry inside the brane is Monkowskian, the AdS geometry of the
exterior region induces vacuum polarization effects in this region as well.
In order to find the corresponding renormalized VEVs of the field square and
the energy-momentum tensor we have presented the Wightman function in the
interior region in decomposed form (\ref{G01}). In this representation the
first term on the right is the Wightman function in the Minkowski spacetime
orbifolded along the direction perpendicular to the brane and the second one
is induced by the AdS geometry in the exterior region. The corresponding
parts in the VEVs of the field square and the energy-momentum tensor are
given by formulae (\ref{phi2intb}), (\ref{T00intb}), (\ref{TDDintb}) and are
investigated in various asymptotic limits of the parameters. For a massless
conformally coupled scalar field these parts vanish. In the general case of
the curvature coupling parameter, the corresponding radial stress is uniform
inside the brane and determine the interior vacuum forces acting on the
boundary of the brane. The results of the numerical calculations plotted in
figures \ref{fig5}, \ref{fig6} show that for both minimally and conformally
coupled scalar fields these forces tend to decrease the thickness of the
brane. When the brane thickness tends to zero, from the formulae of the
model with flat interior the corresponding results in the RS 1-brane model
are obtained.

\section*{Acknowledgments}

The work was supported by the Armenian Ministry of Education and Science
Grant No. 0124.


\begin{thebibliography}{99}
\bibitem{Ruba01} V.A.~Rubakov, Phys. Usp. \textbf{44}, 871 (2001);
R.~Maartens, Living Rev. Relativity \textbf{7}, 7 (2004).

\bibitem{Rand99a} L. Randall and R. Sundrum, Phys. Rev. Lett. \textbf{83},
3370 (1998).

\bibitem{Rand99b} L. Randall and R. Sundrum, Phys. Rev. Lett. \textbf{83},
4690 (1998).

\bibitem{Fabi00} M.~Fabinger and P.~Horava, Nucl. Phys. \textbf{B580}, 243
(2000).

\bibitem{Noji00a} S.~Nojiri, S.D.~Odintsov, and S.~Zerbini, Phys. Rev. D
\textbf{62}, 064006 (2000); S.~Nojiri and S.~Odintsov, Phys. Lett. B \textbf{%
484}, 119 (2000).

\bibitem{Toms00} D.J.~Toms, Phys. Lett. B \textbf{484}, 149 (2000).

\bibitem{Noji00d} S.~Nojiri, O.~Obregon, and S.D.~Odintsov, Phys. Rev. D
\textbf{62}, 104003 (2000).

\bibitem{Gold00} W.~Goldberger and I.~Rothstein, Phys. Lett. B \textbf{491},
339 (2000).

\bibitem{Noji00CQG} S. Nojiri, S.D. Odintsov, and S. Zerbini, Class. Quant.
Grav. \textbf{17}, 4855 (2000); S.~Nojiri and S.~Odintsov, J. High Energy
Phys. \textbf{0007}, 049 (2000).

\bibitem{Garr01} J.~Garriga, O.~Pujolas, and T.~Tanaka, Nucl. Phys. \textbf{%
B605}, 192 (2001).

\bibitem{Muko01} S.~Mukohyama, Phys. Rev. D \textbf{63}, 044008 (2001).

\bibitem{Hofm01} R.~Hofmann, P.~Kanti, and M.~Pospelov, Phys. Rev. D \textbf{%
63}, 124020 (2001).

\bibitem{Brev01} I.~Brevik, K.A.~Milton, S.~Nojiri, and S.D.~Odintsov, Nucl.
Phys. \textbf{B599}, 305 (2001).

\bibitem{Flac01b} A.~Flachi and D.J.~Toms, Nucl. Phys. \textbf{B610}, 144
(2001).

\bibitem{Gilk01} P.B.~Gilkey, K.~Kirsten, and D.V.~Vassilevich, Nucl. Phys.
\textbf{B601}, 125 (2001).

\bibitem{Flac01c} A.~Flachi, I.G.~Moss, and D.J.~Toms, Phys. Lett. B \textbf{%
518}, 153 (2001); Phys. Rev. D \textbf{64}, 105029 (2001).

\bibitem{Nayl02} W.~Naylor and M.~Sasaki, Phys. Lett. B \textbf{542}, 289
(2002).

\bibitem{Saha03} A.A. Saharian and M.R. Setare, Phys. Lett. B \textbf{552},
119 (2003).

\bibitem{Eliz03} E.~Elizalde, S.~Nojiri, S.D.~Odintsov, and S.~Ogushi, Phys.
Rev. D \textbf{67}, 063515 (2003).

\bibitem{Garr03} J.~Garriga and A.~Pomarol, Phys. Lett. B \textbf{560}, 91
(2003).

\bibitem{Noji03} S. Nojiri and S.D. Odintsov, J. Cosmol. Astropart. Phys.
\textbf{06}, 004 (2003).

\bibitem{Yera03} A.H.~Yeranyan and A.A.~Saharian, Astrophysics \textbf{46},
386 (2003).

\bibitem{Moss03} I.G.~Moss, W.~Naylor, W.~Santiago-Germ\'{a}n, and
M.~Sasaki, Phys. Rev. D \textbf{67} 125010 (2003).

\bibitem{Saha04PLB} A.A.~Saharian and M.R.~Setare, Phys. Lett. B \textbf{584}%
, 306 (2004).

\bibitem{Noji04a} G. Cognola, E. Elizalde, S. Nojiri, S. D. Odintsov, and S.
Zerbini, Mod. Phys. Lett. A \textbf{19}, 1435 (2004).

\bibitem{Knap03} A.~Knapman and D.J.~Toms, Phys. Rev. D \textbf{69}, 044023
(2004).

\bibitem{Saha04} A.A. Saharian, Astrophysics \textbf{47}, 303 (2004).

\bibitem{Noji04} S. Nojiri and S.D. Odintsov, Phys. Rev. D \textbf{69},
023511 (2004).

\bibitem{Saha04b} A.A. Saharian and M.R. Setare, Phys. Lett. B \textbf{584},
306 (2004).

\bibitem{Norm04} J.P.~Norman, Phys.Rev. D \textbf{69}, 125015 (2004).

\bibitem{Saha04surf} A.A. Saharian, Phys. Rev. D \textbf{70}, 064026 (2004);
A.A. Saharian, Astrophysics \textbf{48}, 122 (2005).

\bibitem{Pujo04} O.~Pujol\`{a}s and T.~Tanaka, J. Cosmol. Astropart. Phys.
\textbf{12}, 009 (2004).

\bibitem{Seta04} M.R. Setare, Eur. Phys. J. C \textbf{38}, 373 (2004).

\bibitem{Saha05} A.A.~Saharian, Nucl. Phys. \textbf{B712}, 196 (2005).

\bibitem{Nayl05} W.~Naylor and M.~Sasaki, Prog. Theor. Phys. \textbf{113},
535 (2005).

\bibitem{Saha05b} A.A. Saharian and M.R. Setare, Nucl. Phys. \textbf{B724},
406 (2005); A. A. Saharian and M. R. Setare, J. High Energy Phys. \textbf{02}%
, 089 (2007).

\bibitem{Pujo05} O. Pujol\`{a}s and M. Sasaki, J. Cosmol. Astropart. Phys.
\textbf{09}, 002 (2005).

\bibitem{Seta05} M.R. Setare, Phys. Lett. B \textbf{620}, 111 (2005).

\bibitem{Eliz06} E. Elizalde, J. Phys. A \textbf{39}, 6299 (2006).

\bibitem{Mina06a} M. Minamitsuji, W. Naylor, and M. Sasaki, Nucl. Phys.
\textbf{B737}, 121 (2006); M. Minamitsuji, W. Naylor, and M. Sasaki, Phys.
Lett. B \textbf{633}, 607 (2006).

\bibitem{Seta06} M.R. Setare, Phys. Lett. B \textbf{637}, 1 (2006).

\bibitem{Saha06c} A.A. Saharian and M.R. Setare, Phys. Lett. B \textbf{637},
5 (2006).

\bibitem{Flac03b} A.~Flachi, J.~Garriga, O.~Pujol\`{a}s, and T.~ Tanaka, J.
High Energy Phys. \textbf{0308}, 053 (2003); A.~Flachi and O.~Pujol\`{a}s,
Phys. Rev. D \textbf{68}, 025023 (2003); A.A. Saharian, Phys. Rev. D \textbf{%
73}, 044012 (2006); A.A. Saharian, Phys. Rev. D \textbf{73}, 064019 (2006);
A.A. Saharian, Phys. Rev. D \textbf{74}, 124009 (2006).

\bibitem{Dewo00} O. DeWolfe, D.Z. Freedman, S.S. Gubser, and A. Karch, Phys.
Rev. D \textbf{62}, 046008 (2000); M. Gremm, Phys. Lett. B
\textbf{478}, 434 (2000); C. Csaki, J. Erlich, T.J. Hollowood, and
Y. Shirman, Nucl. Phys. \textbf{B581}, 309 (2000); S. Kobayashi,
K. Koyama, and J. Soda, Phys. Rev. D \textbf{65}, 064014 (2002);
P. Mounaix and D. Langlois, Phys. Rev. D \textbf{65}, 103523
(2002); N. Sasakura, J. High Energy Phys. \textbf{0202}, 026
(2002); N. Sasakura, Phys. Rev. D \textbf{66}, 065006 (2002); R.
Mansouri, M. Borhani, and S. Khakshournia, Int. J. Mod. Phys. A
\textbf{19}, 4687 (2004); K.A. Bronnikov and B.E. Meierovich,
Grav. Cosmol. \textbf{9}, 313 (2003); O. Castillo-Felisola, A.
Melfo, N. Pantoja, and A. Ram\'{\i}rez, Phys.Rev. D \textbf{70},
104029 (2004); I. Navarro and J. Santiago, JCAP \textbf{0603}, 015
(2006); N. Barbosa-Candejas and A. Herrera-Aguilar, J. High Energy
Phys. \textbf{0510}, 101 (2005); V. Dzhunushaliev, gr-qc/0603020;
S. Ghassemi, S. Khakshournia, and R. Mansouri, J. High Energy
Phys. \textbf{0608}, 019 (2006);  N. Barbosa-Candejas and A.
Herrera-Aguilar, Phys.Rev. D \textbf{73}, 084022 (2006); R.
Moderski and M. Rogatko, Phys. Rev. D \textbf{74}, 044002 (2006);
S. Ghassemi, S. Khakshournia, and R. Mansouri, gr-qc/0609132.

\bibitem{Birrell} N.D.~Birrell and P.C.W.~Davis, \textit{Quantum Fields in
Curved Space} (Cambridge University Press, 1982).

\bibitem{Brei82} P. Breitenlohner and D.Z. Freedman, Phys. Lett. B \textbf{%
115}, (1982); P. Breitenlohner and D.Z. Freedman, Ann. Phys. (N.Y.) \textbf{%
144}, 249 (1982); L. Mezincescu and P.K. Townsend, Ann.Phys. (N.Y.) \textbf{%
160}, 406 (1985).

\bibitem{Burg85} C.P. Burgess and C.A. L\"{u}tken, Phys. Lett. B \textbf{153}%
, 137 (1985); R. Camporesi, Phys. Rev. D \textbf{43}, 3958 (1991); M. Camela
and C.P. Burgess, Can. J. Phys. \textbf{77}, 85 (1999); M.M. Caldarelli,
Nucl. Phys. \textbf{B549}, 499 (1999).

\bibitem{SahaSurf} A.A. Saharian, Phys. Rev. D \textbf{69}, 085005 (2004).

\bibitem{Prud86} A.P. Prudnikov, Yu.A. Brychkov, and O.I. Marichev, \textit{%
Integrals and series,} vol.2 (Gordon \& Breach, New York, 1986).

\bibitem{Alle90} B. Allen and A.C. Ottewill, Phys. Rev. D \textbf{42}, 2669
(1990); B. Allen, B.S. Kay, and A.C. Ottewill, Phys. Rev. D \textbf{53},
6829 (1996); E.R. Bezerra de Mello, V.B. Bezerra, A.A. Saharian, and A.S.
Tarloyan, Phys. Rev. D \textbf{74}, 025017 (2006); E. R. Bezerra de Mello
and A. A. Saharian, J. High Energy Phys. \textbf{0610}, 049 (2006); E. R.
Bezerra de Mello and A. A. Saharian, Phys. Rev. D \textbf{75}, 065019 (2007).

\bibitem{abramowiz} \textit{Handbook of Mathematical Functions}, edited by
M. Abramowitz and I.A. Stegun (Dover, New York, 1972).
\end{thebibliography}
\end{document}